\documentclass[aip,jcp,reprint,showpacs,preprintnumbers,amsmath,amssymb,twocolumn]{revtex4-1}
\usepackage{graphicx}
\usepackage{dcolumn}
\usepackage{bm}
\usepackage{amssymb}
\usepackage{array}
\usepackage{longtable}
\usepackage{CJK}
\usepackage{color}

\begin{document}


\title{Random walks in unweighted and weighted modular scale-free networks with a perfect trap}

\author{Yihang Yang}

\author{Zhongzhi Zhang}
\email{zhangzz@fudan.edu.cn}
\homepage{http://www.researcherid.com/rid/G-5522-2011}

\affiliation {School of Computer Science, Fudan University,
Shanghai 200433, China}

\affiliation {Shanghai Key Lab of Intelligent Information
Processing, Fudan University, Shanghai 200433, China}

\begin{abstract}

Designing optimal structure favorable to diffusion and effectively controlling the trapping process are crucial in the study of trapping problem---random walks with a single trap. In this paper, we study the trapping problem occurring on unweighted and weighted networks, respectively. The networks under consideration display the striking scale-free, small-world, and modular properties, as observed in diverse real-world systems. For binary networks, we concentrate on three cases of trapping problems with the trap located at a peripheral node, a neighbor of the root with the least connectivity, and a farthest node, respectively. For weighted networks with edge weights controlled by a parameter, we also study three trapping problems, in which the trap is placed separately at the root, a neighbor of the root with the least degree, and a farthest node. For all the trapping problems, we obtain the analytical formulas for the average trapping time (ATT) measuring the efficiency of the trapping process, as well as the leading scaling of ATT. We show that for all the trapping problems in the binary networks with a trap located at different nodes, the dominating scalings of ATT reach the possible minimum scalings, implying that the networks have optimal structure that is advantageous to efficient trapping. Furthermore, we show that for trapping in the weighted networks, the ATT is controlled by the weight parameter, through modifying which, the ATT can behave superlinealy, linearly, sublinearly, or logarithmically with the system size. This work could help improving the design of systems with efficient trapping process and offers new insight into control of trapping in complex systems.
\end{abstract}



\date{\today}
\maketitle


\section{Introduction}

Trapping is a kind of random walks taking place in networks in the presence of a perfect trap, which was introduced in the seminal work by Montroll more than 40 years ago~\cite{Mo69}. As a fundamental dynamical process, it describes or characterizes various phenomena or other dynamical processes in diverse complex systems with frequently cited examples including light harvesting in dendrimeric systems~\cite{BaKlKo97,BaKl98,BaKl98JOL}, page research or access in the World Wide Web~\cite{HwLeKa12,HwLeKa12E}, energy or exciton transport in polymer systems~\cite{SoMaBl97,BlZu81,MuBlAmGiReWe07,AgBlMu10IJBC,Ag11,MuBl11}, and so forth. An essential quantity for trapping problem is trapping time, i.e., mean first-passage time (MFPT)~\cite{Re01,NoRi04,BeCoMo05,CoBeKl07,CoBeMo07,CoBeTeVoKl07}. The trapping time of a node is defined as the expected time for a walker starting off from
this node to visit the trap for the first time. The mean of trapping time to a given trap over all starting nodes is called average trapping time (ATT), which offers useful insight to the trapping process, providing a quantitative measure of trapping efficiency.

One of the major lines of study on trapping has concentrated on understanding how the network topologies affect the behavior of ATT for trapping occurring in different systems. During the past years significant efforts have been devoted to trapping issue in diverse networked systems with particular structural properties, such as square-planar lattices and cubic lattices~\cite{GLKo05,GLLiYoEvKo06}, Sierpinski gasket~\cite{KaBa02PRE,BeTuKo10} and Sierpinski tower~\cite{KaBa02IJBC}, $T-$shape fractal and its extensions~\cite{KaRe89,Ag08,HaRo08,LiWuZh10,ZhWuCh11, WuZh13}, dendrimers~\cite{BeHoKo03,BeKo06,WuLiZhCh12,LiZh13JCP}, hyperbranched polymers~\cite{WuLiZhCh12,LiZh13JCP}, non-fractal~\cite{KiCaHaAr08,ZhQiZhXiGu09,ZhGuXiQiZh09} and fractal scale-free networks~\cite{ZhXiZhGaGu09,ZhXiZhLiGu09,ZhYaGa11}. These works showed that topological properties crucially affect the trapping efficiency measured by ATT, which can display superlinear, linear, sublinear, logarithmical and other dependence on the system size, depending on network structure.

Although ATT in different systems exhibits rich behavior, it has been recently reported~\cite{TeBeVo09,LiJuZh12,LiZh13} that for trapping in any network with a deep trap placed at an arbitrary node, the possible minimal scaling for the ATT to the target is proportional to the network size and the inverse degree of the trap, which is independent of any individual structural parameter of the network. Previous works also provided the mathematical condition under which the maximal scaling for the lower bound of ATT can be reached~\cite{TeBeVo09,LiJuZh12,LiZh13},  however it is difficult to specify existent real or modelling networks in which the predicted minimal scaling of the ATT can be obtained. Therefore, it is of great interest to design or find optimal networks where the minimal scaling of ATT can be achieved.

Another outstanding problem pertaining to trapping is to control the dynamical process~\cite{BaKl98JPC}. Recently, the subject of controlling complex networks towards desired functions has received considerable attention and become an active area of research~\cite{LiSlBa11,YaReLaLaLi12,WaNiLaGr12,LiSlBa13,YuZhDiWaLa13}. In the context of trapping in networks, it is desirable to control the trapping process by using an appropriate approach, with the aim of obtaining needed trapping efficiency. It has been experimentally demonstrated that energy funnel can be applied to modify the trapping efficiency of compact and extended dendrimers~\cite{KoShShTaXuMoBaKl97}, which is actually a control of trapping in polymer networks by changing the local transition probability but keeping the network structure. However, related theoretical analysis on steering trapping process in complex networks, even particular networks towards wanted trapping efficiency is still much less~\cite{BaKl98JPC}.

In this paper, we consider the trapping problem in a family of modular unweighted and weighted scale-free networks~\cite{RaSoMoOlBa02,RaBa03}. For the binary networks, we address three cases of trapping issues with the trap positioned at three representative nodes, i.e., a peripheral node, a neighboring node of the root with the least degree, and a farthest node from the root, respectively. For the weighted networks with the edge weights governed by a tunable parameter, we also address three trapping problems with the immobile trap located at the root node, a root's neighbor with the smallest connectivity, and a farthest node, respectively.

For all the trapping problems occurring in binary and weighted networks, we derive analytically the ATT and their leading scalings. For trapping in unweighted networks, we show that for all cases of trapping problems considered, the possible minimum scalings for ATT can be achieved, implying that the studied networks have optimal structure for trapping with the highest trapping efficiency. For trapping in weighted networks, we show that the ATT exhibits rich behavior, i.e., various dependence on the network size, by tailoring the weight parameter. This work offers instructive clues on designing networks helpful to efficient diffusion and controlling trapping process towards desirable trapping efficiency.

\section{Construction and properties of unweighted and weighted modular scale-free networks}

We first introduce the construction algorithm and structural features of a family of unweighted modular scale-free networks, as well as their weighted counterparts.

\subsection{Unweighted modular scale-free networks}

\begin{figure}
\begin{center}
\includegraphics[width=0.9\linewidth,trim=0 0 0 0]{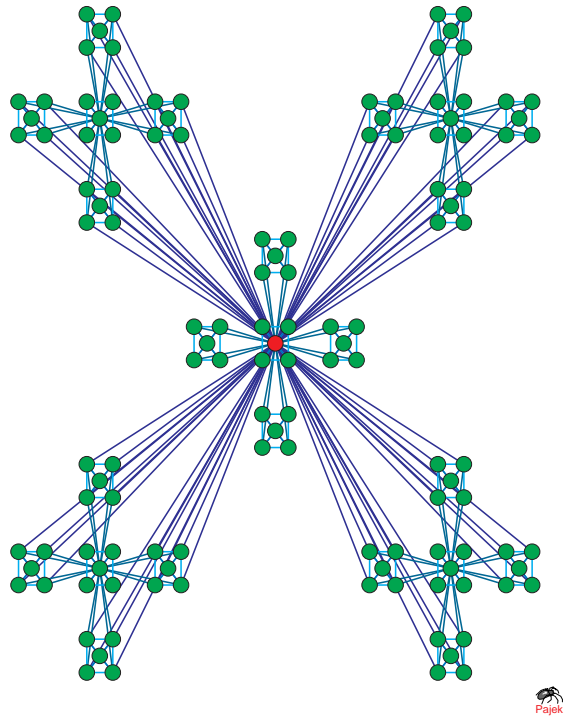}
\end{center}
\caption[kurzform]{(Color online) Structure of network $M_3$ for the
limiting case of $m=5$. Note that the diagonal nodes are also linked to one another; the edges are not visible.} \label{network}
\end{figure}

The family of unweighted modular scale-free networks under consideration is built in an iterative way~\cite{RaSoMoOlBa02,RaBa03}, which is an extension of the hierarchial network
proposed in Ref.~\cite{BaRaVi01} and studied in great detail
in Refs.~\cite{IgYa05,ZhLiGaZhGu09,AgBu09,AgBuMa10,MeAgBeVo12,YaZh13}. Let $M_g$ ($g\geq1$) stand for the networks after $g$ iterations (number of generations). Initially ($g=1$), the network family is composed of a central node, called the root (hub) node, and $m-1$ ($m\geq3$) peripheral nodes. All these $m$ initial nodes are fully connected to each other forming a complete graph. For $g\geq2$, $M_g$ is obtained by adding $m-1$ duplicates, denoted by $M_{g-1}^{(1)}, M_{g-1}^{(2)}, \cdots, M_{g-1}^{(m-1)}$, of $M_{g-1}$ to the original $M_{g-1}$, with all peripheral nodes of the replicas being linked to the root of the primal $M_{g-1}$ unit. In this way, we obtain $M_{g}$, the root and peripheral nodes of which are the root of the original $M_{g-1}$ and the peripheral nodes in the $m-1$ copies of $M_{g-1}$, respectively. Repeating indefinitely the replication and connection steps, we obtain the hierarchical unweighted modular scale-free networks. Figure~\ref{network} illustrates schematically the structure of $M_3$ for the particular case of $m=5$.

According to the above construction algorithm, the number of nodes in $M_g$, denoted as $N_g$, is $N_g=m^g$. All these nodes can be categorized into four different sets~\cite{No03,NoRi04a}: the peripheral node set $\mathbb P$, the locally peripheral node set ${\mathbb P}_z$ ($1\leq z<g$), the set $\mathbb H$ including only the hub node of $M_g$, and the local hub set ${\mathbb H}_z$ ($1\leq z<g$). The cardinalities, defined as the number of nodes in a set, of the four sets are
\begin{equation}\label{A1}
|{\mathbb P}|=(m-1)^g,
\end{equation}
\begin{equation}\label{A2}
|{\mathbb P}_z|=(m-1)^z m^{g-(z+1)},
\end{equation}
\begin{equation}\label{A3}
|{\mathbb H}|=1,
\end{equation}
and
\begin{equation}\label{A4}
|{\mathbb H}_z|=(m-1)m^{g-(z+1)},
\end{equation}
respectively. All nodes belonging to the same set have identical connectivity. The degree for a node in sets $\mathbb H$, ${\mathbb H}_z$, $\mathbb P$, and ${\mathbb P}_z$ is, respectively,
\begin{equation}\label{A5}
K_h(g)=\sum_{g_i=1}^{g}(m-1)^{g_i}=\frac{m-1}{m-2}[(m-1)^g-1],
\end{equation}
\begin{equation}\label{A6}
K_{h,z}(g)=\sum_{g_i=1}^{z}(m-1)^{g_i}=\frac{m-1}{m-2}[(m-1)^z-1],
\end{equation}
\begin{equation}\label{A7}
K_{p}(g)=g+m-2,
\end{equation}
and
\begin{equation}\label{A8}
K_{p,z}(g)=z+m-2.
\end{equation}
Then, the sum of degrees over all nodes in $M_g$ is
\begin{equation}\label{SumDeg}
D_g=(3m-2)(m-1)m^{g-1}-2(m-1)^{g+1}.
\end{equation}
It is easy to check that the networks are sparse with an average degree $2D_g/N_g$, which approximates $2(m-1)(3m-2)/m$ when $N_g$ is very large.

In $M_g$, the maximal value of the shortest distance of all paths from the root to other nodes is $g$. Let $\mathbb F_g$ denote the set of those nodes in $M_g$ at a distance $g$ from the root, hereafter called the farthest nodes of $M_g$. Then, the number of nodes in $\mathbb F_g$ satisfies the relation~\cite{ZhYaLi12}
\begin{equation}\label{A9}
|\mathbb F_g|=(m-1)|\mathbb F_{g-2}|.
\end{equation}
Considering $|\mathbb F_1|=m-1$ and $|\mathbb F_2|=m-1$, the recursive relation can be solved to obtain
\begin{equation}\label{A10}
|\mathbb F_g|=\left\{\begin{aligned}
&(m-1)^{(g+1)/2}, \quad g\,\rm{is\, odd}\\
&(m-1)^{(g/2)}, \quad g\, \rm{is\, even.}\end{aligned}\right.
\end{equation}

The networks being studied present some typical features observed in a variety of real systems. They are power law~\cite{BaAl99} with the degree distribution exponent $\gamma$ being equal to $1+\ln{m}/\ln{(m-1)}$. In addition, they exhibit the small-world effect~\cite{WaSt98}, with small average distance that grows logarithmically with the network size~\cite{No03,ZhLi10} and high clustering coefficient~\cite{RaBa03,No03}. In particular, the networks display the remarkable modular structure~\cite{GiNe02,PaDeFaVi05,Ne06,Fo10} that is observed in various real-life networks, e.g., biological networks and social networks.

\subsection{Weighted modular scale-free networks}

The aforementioned unweighted modular scale-free networks can be extended to weighed networks, by introducing a weight parameter $\omega$ ($\omega>0$) in the construction algorithm. Let $\bar{M}_g$ denote the weighted networks after $g$ generations, which are constructed as follows. For $g=1$, $\bar{M}_1$ is composed of $m$ ($m\geq3$) nodes, of which one is the root node, while the other $m-1$ nodes are peripheral nodes. These $m$ nodes are linked by $m(m-1)/2$ weighted edges forming a complete graph. The weight of any edge linking the root and a peripheral node is equal to $\omega$; while any other edge between an arbitrary pair of peripheral nodes has unit weight. For $g>1$, $\bar{M}_g$ can be obtained by adding $m-1$ copies of $\bar{M}_{g-1}$ to the primal $\bar{M}_{g-1}$, with all peripheral nodes of the replicas being linked to the root of the original $\bar{M}_{g-1}$ unit by $(m-1)^g$ edges, each having identical weight $\omega^{g}$.


In a weighted network, the strength of any node $i$ is defined by~\cite{BaBaVe04}
\begin{equation}\label{A11}
s_i=\sum_{j=1}^N w_{ij},
\end{equation}
where $w_{ij}$ is the weight of the edge linking nodes $i$ and $j$. For the weighted modular networks $\bar{M}_g$, it is easy to derive that the strengths of nodes in ${\mathbb H}$, ${\mathbb H}_z$, ${\mathbb P}$, and ${\mathbb P}_z$, are
\begin{eqnarray}\label{A12}
S_{h}(g)&=&\sum_{g_i=1}^{g}(m-1)^{g_i}\omega^{g_i}\nonumber\\
&=&\frac{(m-1)\omega}{(m-1)\omega-1}[(m-1)\omega]^g-1-\frac{1}{m\omega-\omega-1},\nonumber\\
\end{eqnarray}
\begin{eqnarray}\label{A13}
S_{h,z}(g)&=&\sum_{g_i=1}^{z}(m-1)^{g_i}\omega^{g_i}\nonumber\\
&=&\frac{(m-1)\omega}{(m-1)\omega-1}[(m-1)\omega]^z-1-\frac{1}{m\omega-\omega-1},\nonumber\\
\end{eqnarray}
\begin{eqnarray}\label{A14}
S_{p}(g)=\sum_{g_i=1}^{g}\omega^{g_i}+(m-2)=\frac{\omega}{\omega-1}(\omega^g-1)+m-2,\nonumber\\
\end{eqnarray}
and
\begin{eqnarray}\label{A15}
S_{p,z}(g)=\sum_{g_i=1}^{z}\omega^{g_i}+(m-2)=\frac{\omega}{\omega-1}(\omega^z-1)+m-2,\nonumber\\
\end{eqnarray}
respectively. Thus, the sum of strengths over all nodes in $\bar{M}_g$ is
\begin{small}
\begin{eqnarray}\label{A16}
S&=&S_{h}(g)|\mathbb H|+\sum_{z=1}^{g-1}S_{h,z}(g)|\mathbb H_z|+S_{p}(g)|\mathbb P|+\sum_{z=1}^{g-1}S_{p,z}(g)|\mathbb P_z|\nonumber\\
&=&\frac{2\omega(m-1)}{m\omega-m-\omega}[(m-1)\omega]^g\nonumber\\
&\quad&+\frac{m-1}{m(m\omega-m-\omega)}[(\omega-1)m^2+(2-5\omega)m+2\omega]m^g.\nonumber\\
\end{eqnarray}
\end{small}
For the case of $\omega=1$, the networks become binary, and Eqs.~(\ref{A12}),~(\ref{A13}),~(\ref{A14}),~(\ref{A15}),~(\ref{A16}) reduce to Eqs.~(\ref{A5}),~(\ref{A6}),~(\ref{A7}),~(\ref{A8}), and~(\ref{SumDeg}), respectively. Furthermore, for general $\omega$, as will be shown below, it acts as the similar role of energetic funnel superimposed on the dendrimers~\cite{BaKlKo97,BaKl98,BaKl98JOL}, which introduces the possibility for controlling the efficiency of trapping in the weighted networks $\bar{M}_g$.

After introducing the construction and properties of the unweighted and weighted modular scale-free networks, in the sequel, we will study analytically the trapping process performing on the networks with a perfect trap fixed at a certain node, in order to uncover the impacts of structure and weight on the trapping efficiency.


\section{Trapping in unweighted modular scale-free networks}

The peculiar architecture of the networks makes it worthwhile to study dynamical processes performing on them. In this section we consider discrete unbiased (isotropic) random walks in binary modular scale-free networks $M_g$ with a single trap fixed at a given node. Let $T_{ij}(g)$ denote the MFPT from node $i$ to $j$ in $M_g$, which is the expected time taken by a walker starting from $i$ to first arrive at $j$. The highly desirable quantity related to the trapping problem is the ATT. If node $j$ is the trap, the ATT to $j$, denoted by $T_j(g)$, is defined as the average of $T_{ij}(g)$ over all the $N_g$ source nodes in $M_g$. By definition, $T_j(g)$ is given by
\begin{equation}\label{Z1}
T_j(g)=\frac{1}{N_g}\sum_{i=1}^{N_g} T_{ij}(g).
\end{equation}
Below we will determine explicitly $T_j(g)$ for three cases of trapping problem performed on $M_g$, with the perfect trap placed at a peripheral node, a neighboring node of the root with the least degree, and a farthest node, respectively. For these three representative trapping problems, we will show how the dominating behavior of ATT scales with the network size, so as to extract  information about the intrinsic impacts of network structure on trapping.

\subsection{Trapping with the trap positioned at a peripheral node}

We here consider the case that the trap is positioned at one of the $(m-1)^g$ peripheral nodes. Notice that for this case, the ATT to any peripheral node is identical. Thus, we only focus on a particular trapping problem with the trap located at a given peripheral node. For this purpose,
we first determine the MFPT from the root to the trap, based on which we derive the ATT to the trap in $M_g$.

\subsubsection{Related definitions and quantities}

Prior to deducing the ATT to a peripheral node, we introduce some related quantities. Let $T_{p h}(g)$ and $T_{h p}(g)$ separately denote the MFPT from an arbitrary peripheral node to the hub node of $M_g$ and the MFPT from the hub node to any of $(m-1)^g$ arbitrary peripheral nodes in $M_g$. The two quantities have been derived previously in different approaches~\cite{ZhLiGoZhGuLi09,ZhYaLi12}:
\begin{equation}\label{B1}
T_{p h}(g)=\left(3m-8+\frac{7m-2}{m^2}\right)\left(\frac{m}{m-1}\right)^g-2m+3
\end{equation}
and
\begin{equation}\label{B2}
T_{h p}(g)=\left(3-\frac{5m-2}{m^2}\right)\left(\frac{m}{m-1}\right)^g-1,
\end{equation}
which are very useful for the following derivations. Note that in very large networks, both $T_{h p}(g)$ and $T_{p h}(g)$ scale sublinearly with the network size $N_g$ as $(N_g)^{1-\ln (m-1)/\ln m}$.

In order to determine the ATT to a peripheral node, we further classify all the $(m-1)^g$  peripheral nodes in ${\mathbb P}$ in the following way. First, we label the $(m-1)^g$ peripheral nodes sequentially by $1$, $2$, $\cdots$, $(m-1)^g-1$, and $(m-1)^g$. Then, these $(m-1)^g$ peripheral nodes can be classified into $g+1$ sets denoted by $\beta_{i}$ ($0\leq i\leq g$). For $i=0$, $\beta_0={1}$; while for $1\leq i\leq g$, $\beta_i=\{x|(m-1)^{i-1}<x\leq(m-1)^i\}$. In addition, let ${\mathcal B}_i$ ($0\leq i\leq g$) be the union of the sets $\beta_k$ with $0\leq k\leq i$, namely ${\mathcal B}_i=\bigcup_{k=0}^{i}\beta_k$. Without loss of generality, we can choose the node belonging to $\beta_0={\mathcal B}_0$ as the trap. Thus, $T_{h \mathcal B_0}(g)$ denote the MFPT from the root to the trap, and $T_{{\mathcal B}_0}(g)$ stands for the ATT.

Before evaluating $T_{h \mathcal B_0}(g)$ and $T_{\mathcal B_0}(g)$, we need to define and determine some new quantities. Let $T_{\beta_{i+1}\mathcal B_i}(g)$ be the MFPT for a particle leaving from an arbitrary node in $\beta_{i+1}$ to an arbitrary node belonging to $\mathcal B_i$ in $M_g$. To determine $T_{\beta_{i+1} \mathcal B_i}(g)$, we distinguish two cases: $i>0$ and $i=0$. For the case $i>0$, $T_{\beta_{i+1} \mathcal B_i}(g)$ satisfies the relation
\begin{small}
\begin{eqnarray}\label{B3}
T_{\beta_{i+1} \mathcal B_i}(g)&=&\frac{1}{g+m-2}\Bigg[(m-2)(1+T_{\beta_{i+1} \mathcal B_i}(g))\nonumber\\
&\quad&+\sum_{k=1}^{i}\left(1+T_{hp}(k)+T_{\beta_{i+1} \mathcal B_i}(g)\right)\nonumber\\
&\quad&+\sum_{k=i+1}^{g}\left(1+T_{hp}(k)+\frac{m-2}{m-1}\sum_{l=i}^{k-1}T_{\beta_{l+1} \mathcal B_{l}}(g)\right)\Bigg].\nonumber\\
\end{eqnarray}
\end{small}
The three terms on the right-hand side (rhs) of Eq.~(\ref{B3}) can be explained as follows. The first term is based on the fact that the walker takes one time step to reach another peripheral node in $\beta_{i+1}$ and then jumps $T_{\beta_{i+1} \mathcal B_i}(g)$ more steps to reach the target node for the first time. The second term describes the process by which the particle first jumps to a local hub node that has no links to other peripheral nodes except those in $\beta_{i+1}$, then makes $T_{hp}(k)+T_{\beta_{i+1} \mathcal B_i}(g)$ jumps to the trap. The last term accounts for the fact that the walker first hits a local hub that has a link connected to peripheral nodes not in $\beta_{i+1}$, then takes $T_{h p}(k)+\frac{m-2}{m-1}\sum_{l=i}^{k-1}T_{\beta_{l+1} \mathcal B_{l}}(g)$ steps to visit the destination.

According to Eq.~(\ref{B3}), we have
\begin{eqnarray}\label{B4}
T_{\beta_{i+1} \mathcal B_i}(g)&=&\frac{m-1}{g-i}\bigg(g+m-2+\sum_{k=1}^{g}T_{hp}(k)\nonumber\\
&\quad&+\frac{m-2}{m-1}\sum_{k=i+1}^{g}\sum_{l=i+1}^{k-1}T_{\beta_{l+1} \mathcal B_l}(g)\bigg)
\end{eqnarray}
and
\begin{eqnarray}\label{B5}
T_{\beta_{i} \mathcal B_{i-1}}(g)&=&\frac{m-1}{g-i+1}\bigg(g+m-2+\sum_{k=1}^{g}T_{hp}(k)\nonumber\\
&\quad&+\frac{m-2}{m-1}\sum_{k=i}^{g}\sum_{l=i}^{k-1}T_{\beta_{l+1} \mathcal B_{l}}(g)\bigg),
\end{eqnarray}
both of which give rise to
\begin{eqnarray}\label{B6}
T_{\beta_i \mathcal B_{i-1}}(g)=\frac{(m-1)(g-i)}{g-i+1}T_{\beta_{i+1} \mathcal B_i}(g).
\end{eqnarray}
It is easy to derive that
\begin{eqnarray}\label{B7}
T_{\beta_{g} \mathcal B_{g-1}}(g)=(m-1)\left(g+m-2+\sum_{k=1}^{g}T_{hp}(k)\right),
\end{eqnarray}
which, together with Eq.~(\ref{B2}) yields
\begin{eqnarray}\label{B8}
T_{\beta_{g} \mathcal B_{g-1}}(g)=\frac{(m-1)^2}{m}\left[(3m-2)\left(\frac{m}{m-1}\right)^{g}-2m\right].
\end{eqnarray}
Considering the initial condition in Eq.~(\ref{B8}), Eq.~(\ref{B6}) can be solved to obtain
\begin{eqnarray}\label{B9}
T_{\beta_{i+1} \mathcal B_i}(g)=\frac{(m-1)^{g-i+1}}{(g-i)m}\left[(3m-2)\left(\frac{m}{m-1}\right)^g-2m\right].
\end{eqnarray}

For the case $i=0$, the $T_{\beta_{1} \mathcal B_0}(g)$ is given by
\begin{eqnarray}\label{B10}
T_{\beta_{1} \mathcal B_0}(g)&=&\frac{1}{g+m-2}\Bigg[1+(m-3)(1+T_{\beta_1 \mathcal B_0}(g))\nonumber\\
&\quad&+\sum_{k=1}^{g}\left(1+T_{hp}(k)+\frac{m-2}{m-1}\sum_{l=0}^{k-1}T_{\beta_{l+1} \mathcal B_l}(g)\right)\Bigg].\nonumber\\
\end{eqnarray}
Note that in this case the walker may directly take one step to arrive at the trap. On the other hand, from Eq.~(\ref{B4}) one has
\begin{eqnarray}\label{B11}
T_{\beta_{2} \mathcal B_1}(g)&=&\frac{m-1}{g-1}\bigg[g+m-2+\sum_{k=1}^{g}T_{hp}(k)\nonumber\\
&\quad&+\frac{m-2}{m-1}\sum_{k=2}^{g}\sum_{l=2}^{k-1}T_{\beta_{l+1} \mathcal B_l}(g)\bigg]\,.
\end{eqnarray}
From the above two equations, we can obtain the following recursion relation governing $T_{\beta_1 \mathcal B_0}(g)$ and $T_{\beta_{2} \mathcal B_1}(g)$:
\begin{eqnarray}\label{B12}
T_{\beta_1 \mathcal B_0}(g)=\frac{(g-1)(m-1)}{g+m-1}T_{\beta_2 \mathcal B_1}(g).
\end{eqnarray}
Inserting the value of $T_{\beta_2 \mathcal B_1}$ provided in Eq.~(\ref{B9}) into Eq.~(\ref{B12}) leads to
\begin{eqnarray}\label{B13}
T_{\beta_1 \mathcal B_0}(g)=\frac{(m-1)^{g+1}}{m(g+m-1)}\left[(3m-2)\left(\frac{m}{m-1}\right)^g-2m\right].
\end{eqnarray}

After obtaining the expressions of related quantities, we next determine the MFPT $T_{h \mathcal B_0}(g)$ from the hub node to the trap, as well as the ATT $T_{\mathcal B_0}(g)$.

\subsubsection{MFPT from the root to the trap}

The above obtained intermediate quantities enable us to evaluate $T_{h \mathcal B_0}(g)$. For random walks in $M_g$, let $T_{h \mathcal B_i}(g)$ be the MFPT from the root to an arbitrary node belonging to $\mathcal B_i$, which follows the relation:
\begin{equation}\label{C1}
T_{h \mathcal B_i}(g)=T_{h \mathcal B_{i+1}}(g)+\left(1-\frac{1}{m-1}\right)T_{\beta_{i+1} \mathcal B_{i}}(g).
\end{equation}
Equation~(\ref{C1}) can be elaborated as follows. For a walker starting from the root, in order to reach nodes in $\mathcal B_i$, it must first take $T_{h \mathcal B_{i+1}}(g)$ time steps to arrive at a node in $\mathcal B_{i+1}$, among which the proportion of nodes belonging to $\mathcal B_i$ is $1/(m-1)$. If the walker first makes a jump to other nodes not in $\mathcal B_i$, it should jump more $T_{\beta_{i+1} \mathcal B_i}(g)$ to reach an arbitrary target node in $\mathcal B_i$, a process happening with a complementary probability of $1/(m-1)$.

By the definition of $T_{h \mathcal B_i}(g)$, it is easy to get the initial condition of Eq.~(\ref{C1}): $T_{h \mathcal B_g}(g)=T_{h p}(g)$. Combining Eqs.~(\ref{B2}) and~(\ref{B9}), Eq.~(\ref{C1}) is solved to yield
\begin{small}
\begin{eqnarray}\label{C2}
T_{h \mathcal B_i}(g)&=&\frac{m-2}{m}\left[(3m-2)\left(\frac{m}{m-1}\right)^g-2m\right]\sum_{k=1}^{g-i}\frac{(m-1)^k}{k}\nonumber\\
&\quad&+\left(3-\frac{5m-2}{m^2}\right)\left(\frac{m}{m-1}\right)^g-1\,,
\end{eqnarray}
\end{small}
which holds for $i>0$.

While for $i=0$, according to Eq.~(\ref{C1}) we have
\begin{eqnarray}\label{C3}
T_{h \mathcal B_0}(g)=T_{h \mathcal B_1}(g)+\frac{m-2}{m-1}T_{\beta_1 \mathcal B_0}(g).
\end{eqnarray}
Instituting Eqs.~(\ref{B13}) and (\ref{C2}) into Eq.~(\ref{C3}), the closed-form expression for MFPT from the hub node to the trap is given by
\begin{small}
\begin{eqnarray}\label{C4}
T_{h \mathcal B_0}(g)&=&\frac{m-2}{m}\left[(3m-2)\left(\frac{m}{m-1}\right)^g-2m\right]\sum_{k=1}^{g-1}\frac{(m-1)^k}{k}\nonumber\\
&\quad&+\left(3-\frac{5m-2}{m^2}\right)\left(\frac{m}{m-1}\right)^g-1\nonumber\\
&\quad&+\frac{(m-2)(m-1)^g}{m(g+m-1)}\left[(3m-2)\left(\frac{m}{m-1}\right)^g-2m\right].\nonumber\\
\end{eqnarray}
\end{small}
It is not difficult to find that the term with the highest exponent occurs when $k=g-1$. Moreover, in the infinite network size limit, i.e., $N_g\to\infty$, we have
\begin{eqnarray}\label{C5}
T_{h \mathcal B_0}(g)\sim m^g/g =N_g/ \ln N_g,
\end{eqnarray}
that is, the leading term of $T_{h \mathcal B_0}(g)$ grows linearly with the network size by a logarithmical correction.

\subsubsection{Exact solution and dominating scaling for ATT}

By construction of $M_g$, the ATT to the trap can be evaluated as follows:
\begin{widetext}
\begin{eqnarray}\label{B14}
T_{\mathcal B_0}(g)&=&\frac{1}{m^g}\left[T_{hp}(1)+\frac{m-2}{m-1}T_{\beta_{1} \mathcal B_0}(g)+(m-2)T_{\beta_1 \mathcal B_0}(g)\right]+\sum_{i=2}^{g}\frac{1}{m^{g+1-i}}\Bigg[T_{h}(i-1)+T_{hp}(i)+\frac{m-2}{m-1}\sum_{k=0}^{i-1}T_{\beta_{k+1}, \mathcal B_k}(g)\nonumber\\
&\quad&+(m-2)\left(T_{p}(i-1)+T_{\beta_{i} \mathcal B_{i-1}}(g)+\frac{m-2}{m-1}\sum_{k=0}^{i-2}T_{\beta_{k+1} \mathcal B_k}(g)\right)\Bigg],\nonumber\\
\end{eqnarray}
\end{widetext}
where $T_{h}(g)$ is the ATT to the root and $T_{p}(g)$ is the ATT when all peripheral nodes are occupied by traps, both of which have been studied in Ref.~\cite{ZhYaLi12} and are given by
\begin{eqnarray}\label{B15}
T_{h}(g)&=&\frac{2(3m-2)(m-1)^3}{m^3}\left(\frac{m}{m-1}\right)^g\nonumber\\
&\quad&-\frac{2(m-1)^2}{m^2}g-\frac{m-1}{m^2}(5m^2-10m+4)
\end{eqnarray}
and
\begin{eqnarray}\label{B16}
T_{p}(g)&=&\frac{(m-1)(3m-2)(m^2-2m+2)}{m^3}\left(\frac{m}{m-1}\right)^g\nonumber\\
&\quad&-\frac{2(m-1)^2}{m^2}g-3m+10-\frac{12}{m}+\frac{4}{m^2}\,,
\end{eqnarray}
respectively.

Although the analytical expression for the ATT $T_{\mathcal B_0}(g)$ to a peripheral node is rather lengthy and awkward, it is easy to infer that when $g$ is large enough, the dominant term of $T_{\mathcal B_0}(g)$ is identical to that of $T_{h \mathcal B_0}(g)$, which can be understood from the following heuristic arguments. As shown above, $M_g$ consists of $m$ subgraphs, which are replicas of $M_{g-1}$. For those nodes in the central subgraph, their ATT to the trap is equal to $T_{h}(g-1)+T_{h \mathcal B_0}(g)$, the dominant term of which is $T_{h \mathcal B_0}(g)$; for nodes in each of the $m-2$ fringe subgraphs $M_{g-1}^{(x)}$ ($x=2,3,\ldots, m-1$) excluding the trap, their MFPT to the trap is $T_{p}(g)+T_{ph}(g)+T_{h \mathcal B_0}(g)$, whose leading term is $T_{h \mathcal B_0}(g)$; while for for nodes in the fringe subgraph $M_{g-1}^{(1)}$ containing the trap, their ATT to the trap is smaller than $T_{p}(g)+T_{ph}(g)+T_{h \mathcal B_0}(g)$.

Hence, for trapping in $M_g$ with a trap placed at a peripheral node, the dominating term of the ATT $T_{\mathcal B_0}(g)$ is the same as $T_{h \mathcal B_0}(g)$ but its prefactor may be different from that of $T_{h \mathcal B_0}(g)$. That is to say, the ATT $T_{\mathcal B_0}(g)$ scales with the network size as
\begin{eqnarray}\label{B17}
T_{\mathcal B_0}(g)\sim N_g/ \ln N_g\,.
\end{eqnarray}

\subsection{Trapping with the trap located at a neighbor of the root with the least degree}

We now address random walks in $M_g$ with a trap located at one of the $m-1$ neighbors of the root, which are local peripheral nodes in $\mathbb P_1$. Let $\Omega$ be the set of these $m-1$ nodes, which are equivalent to one another in the sense that their ATT is identical. Without loss of generality, we choose an arbitrary node in $\Omega$ as the trap and label it by $x$. What we are concerned with is the ATT to the trap node $x$ for trapping in $M_g$, denoted by $T_x(g)$. For the sake of evaluating $T_x(g)$, we first determine the MFPT $T_{h x}(g)$ from the root to the trap $x$. According to the structure of $M_g$, we can establish the following relation:
\begin{eqnarray}\label{D1}
T_{h x}(g)&=&\frac{1}{\sum_{i=1}^g (m-1)^i}\bigg[1+(m-2)(1+T_{y x}(g))\nonumber\\
&\quad&+\sum_{i=2}^{g}(m-1)^i(1+T_{ph}(i)+T_{hx}(g))\bigg],
\end{eqnarray}
where $T_{yx}(g)$ is the MFPT from a node $y$ ($y \neq x$) in $\Omega$ to the trap node $x$ and follows the relation
\begin{eqnarray}\label{D2}
T_{y x}(g)=\frac{1}{m-1}\left[1+(m-3)(1+T_{yx}(g))+(1+T_{hx}(g))\right].\nonumber\\
\end{eqnarray}

The three terms in the square brackets on the rhs of Eq.~(\ref{D1}) can be accounted for as follows. The first term describes the process that the walker originating from the root goes directly to the trap node $x$. The second term presents the process that the particle first jumps to one of the $m-2$ non-trap nodes, say $y$, in $\Omega$ and then takes time $T_{y x}(g)$ to reach the target node. The last sum term explains the fact that the particle goes to a local peripheral node in $\mathbb P_i$ ($2\leq i\leq g$), from which it takes time $T_{p h}(i)$ to return the root, and then takes $T_{h x}(g)$ steps to get to the destination. Analogously, we can explain Eq.~(\ref{D2}).

After some algebra, Eq.(\ref{D2}) can be simplified to
\begin{eqnarray}\label{D3}
T_{y x}(g)=\frac{1}{2}(m-1+T_{h x}(g)),
\end{eqnarray}
inserting which into Eq.~(\ref{D1}) yields
\begin{eqnarray}\label{D4}
T_{h x}(g)&=&\frac{2}{m}\bigg[\sum_{i=2}^g(m-1)^iT_{p  h}(i)+\sum_{i=1}^g(m-1)^i\nonumber\\
&\quad&+\frac{1}{2}(m-1)(m-2)\bigg].
\end{eqnarray}
Plugging Eq.~(\ref{B1}) into Eq.~(\ref{D4}), the MFPT from the hub to the trap can is given by
\begin{eqnarray}\label{D5}
T_{h x}(g)=(6m-4)(m-1)m^{g-2}-\frac{4}{m}(m-1)^{g+1}-m+1.\nonumber\\
\end{eqnarray}

Using the obtained expression for $T_{h x}(g)$, the quantity $T_{x}(g)$ can be accurately evaluated as
\begin{eqnarray}\label{D6}
T_x(g)&=&T_h(g)+T_{h x}(g)-\frac{1}{m^g}\left[\frac{1}{2}m(m-1)+\frac{m}{2}T_{h x}(g)\right]\nonumber\\
&=&2(m-1)(3m-2)m^{g-2}-\frac{4}{m}(m-1)^{g+1}\nonumber\\
&\quad&+2(3m-2)\left(\frac{m}{m-1}\right)^{g-3}+2(m-1)\left(\frac{m-1}{m}\right)^g\nonumber\\
&\quad&-\frac{1}{m^2}[(m-1)(3m-2)^2+2g(m-1)^2],
\end{eqnarray}
which can be expressed in terms of network size $N_g$ as
\begin{small}
\begin{eqnarray}\label{D7}
T_x(g)&=&\frac{2(m-1)(3m-2)}{m^2}N_g-\frac{4(m-1)}{m}(N_g)^{\log_{m}(m-1)}\nonumber\\
&\quad&\frac{2(3m-2)m^3}{(m-1)^3}(N_g)^{\log_{m}{\frac{m}{m-1}}}+2(m-1)(N_g)^{\log_{m}{\frac{m-1 }{m}}}\nonumber\\
&\quad&-\frac{1}{m^2}[(m-1)(3m-2)^2+2g(m-1)^2].
\end{eqnarray}
\end{small}
Thus, when $N_g\to\infty$,
\begin{eqnarray}\label{D8}
T_x(g)\sim N_g,
\end{eqnarray}
implying that $T_x(g)$ grows linearly with the network size.

\subsection{Trapping with the trap fixed a farthest node}

Now we address the problem with the trap being positioned at one of the $|\mathbb F|_g$ farthest nodes in $M_g$. We use $T_{f}(g)$ to denote the ATT to the trap. In order to determine the behavior of $T_{f}(g)$, we first focus on the MFPT from the root to the trap, denoted by $T_{h f}(g)$, based on which we will further show that the leading scaling of $T_{f}(g)$ is identical to that of $T_{h f}(g)$.

By construction, $M_g$ is composed of a primal (central) $M_{g-1}$ and $m-1$ copies of $M_{g-1}$, denoted separately by $M_{g-1}^{(x)}$ ($x=1,2,\ldots, m-1$). In $M_1$, the farthest nodes are exactly its $m-1$ peripheral nodes, and in $M_2$, the farthest nodes correspond to the hub nodes of all $M_{1}^{(x)}$. And in $M_g$ ($g \geq 3$), the $|\mathbb F|_g$ farthest nodes belong to all subgraphs $M_{g-1}^{(x)}$, that is, the farthest nodes of the primal central subgraphs (i.e., $M_{g-2}$) forming $M_{g-1}^{(x)}$. Since for this trapping problem, the ATT to any farthest node is the same, we select a farthest node in $M_{g-1}^{(1)}$ as the deep trap.

\subsubsection{Determination of intermediate variables}

In order to evaluate $T_{h f}(g)$, we introduce some more intermediate quantities. For the nodes in $M_{g-1}^{(x)}$ that are components of $M_g$, we can classify them in the following way. Let $\mathcal H_{g-i}$ ($0\leq i\leq g-1$) be the set of local hub nodes which are directly linked to $g-i$ classes of local peripheral nodes belonging to $\mathbb P_{k}$, and let $\mathcal P_{g-i}$ ($0\leq i\leq g-1$) denote the set of local peripheral nodes that connect to $g-i$ different local hub nodes in $\mathbb H_{k}$. Moreover, we assume that $\mathcal H_g=\mathbb H$ and $\mathcal P_g=\mathbb P$.

For a particle starting from the root to visit one of the $|\mathbb F_g|$ farthest nodes, it must follow the walking path $\mathcal H_g\to\mathcal P_g\to\mathcal H_{g-1}\to\mathcal P_{g-2}\to\mathcal H_{g-3}\to\cdots\to\mathcal P_{g-(i-1)}\to\mathcal H_{g-i}\to\mathcal P_{g-(i+1)}\to\mathcal H_{g-(i+2)}\to\cdots\to\mathcal H_1$ (or $\mathcal P_1$). For the special  case that a farthest node in $M_{g-1}^{(1)}$ is considered as the trap, the path should be definitely as follows: each time the particle starting from a current local hub belonging to $\mathcal H_{g-i}$ in $M_{g-1}^{(1)}$, it must jump to a local peripheral node in $\mathcal P_{g-(i+1)}$, then continues to hop towards a main hub of a subgraph $M_{g-(i+2)}$ that is in the central a subgraph $M_{g-i}$. In this way, the walker moves on until it reaches the trap.

According to the above analysis, for the purpose to determine $T_{h f}(g)$, it is necessary to define two more variables $p_g(i)$ and $h_g(i)$, where the former is the MFPT from a node in $\mathcal P_{g-i}$ to any of its neighbors that simultaneously belongs to $\mathcal H_{g-(i+1)}$, and the latter is the MFPT from a node in $\mathcal H_{g-i}$ to any of its adjacent nodes in both $\mathcal P_{g-(i+1)}$ and $M_{g-(i+2)}^{(1)}$. In Appendix~\ref{Ap1}, we provide the detailed derivation for $p_g(i)$ and $h_g(i)$. For $i<g-2$, we have
\begin{small}
\begin{eqnarray}\label{E1}
p_g(i)&=&(3m-2)\frac{m^{g-1}}{(m-1)^{g-i-2}}-\frac{3m-2}{m}\left(\frac{m}{m-1}\right)^{g-i-2}\nonumber\\
&\quad&-2(m-1)^{i+2}+1
\end{eqnarray}
\end{small}
and
\begin{small}
\begin{eqnarray}\label{E2}
h_g(i)&=&(3m-2)\frac{m^{g-1}}{(m-1)^{g-i-3}}-(3m-2)\left(\frac{m}{m-1}\right)^{g-i-3}\nonumber\\
&\quad&-2(m-1)^{i+3}+2m-3\,.
\end{eqnarray}
\end{small}
For $i=g-2$, $p_g(i)$ and $h_g(i)$ read
\begin{equation}\label{E3}
p_g(g-2)=(3m-2)m^{g-1}-2(m-1)^g-1
\end{equation}
and
\begin{small}
\begin{eqnarray}\label{E4}
h_g(g-2)&=&2(3m-2)(m-1)m^{g-2}-\frac{2}{m}(3m-2)(m-1)\nonumber\\
&\quad&-\frac{4}{m}(m-1)^{g+1}+\frac{1}{m}(5m-4)(m-1)\,,
\end{eqnarray}
\end{small}
respectively.

\subsubsection{Exact solution and leading scaling of MFPT from the root to the trap}

Using the above obtained expressions for $p_g(i)$ and $h_g(i)$, we can derive the formulas for the MFPT $T_{h f}(g)$. In order to get the explicit formula for $T_{h f}(g)$, we distinguish two cases: (i) $g$ is odd and (ii) $g$ is even.

When $g$ is odd, the target node belongs to $\mathcal P_1$. For this case, one has
\begin{eqnarray}\label{F1}
T_{h f}(g)&=&\left[T_{h p}(g)+\frac{m-2}{m-1}T_{\beta_g \mathcal B_{g-1}}(g)\right]+\sum_{i=0}^{\frac{g-1}{2}-1}p_g(2i)\nonumber\\
&\quad&+\sum_{i=0}^{\frac{g-1}{2}-2}h_g(2i+1)+h_g(g-2)\,.
\end{eqnarray}
Substituting Eqs.~(\ref{B2}),~(\ref{E1}),~(\ref{E2}), and~(\ref{E4}) into Eq.~(\ref{F1}), after some algebra, Eq.~(\ref{F1}) is solved to yield the exact solution to $T_{h f}(g)$, given by
\begin{small}
\begin{eqnarray}\label{F2}
T_{h f}(g)&=&\frac{2(3m-2)(m-1)^2}{m-2}m^{g-2}-\frac{4}{m(m-2)}(m-1)^{g+2}\nonumber\\
&\quad&-\frac{(3m-2)(m-1)(m^2+2)}{(m-2)(2m-1)}\left(\frac{m}{m-1}\right)^{g-2}\nonumber\\
&\quad&+\frac{m-1}{m(m-2)(2m-1)}[(2m^3-5m^2+2m)g\nonumber\\
&\quad&+3m^4-9m^3+14m^2-12m+4].
\end{eqnarray}
\end{small}

When $g$ is even, the trap is $\mathcal H_1$. In this case, $T_{h f}(g)$ can be derived by
\begin{eqnarray}\label{F3}
T_{h f}(g)&=&\left[T_{h,p}(g)+\frac{m-2}{m-1}T_{\beta_{g} \mathcal B_{g-1}}(g)\right]+\sum_{i=0}^{\frac{g}{2}-2}p_g(2i)\nonumber\\
&\quad&+\sum_{i=0}^{\frac{g}{2}-2}h_g(2i)+p_g(g-2).
\end{eqnarray}
Plugging Eqs.~(\ref{B2}),~(\ref{E1}),~(\ref{E2}), and~(\ref{E3}) into Eq.~(\ref{F3}), we obtain a closed-form solution to $T_{h f}(g)$ as
\begin{small}
\begin{eqnarray}\label{F4}
T_{h f}(g)&=&\frac{2(3m-2)(m-1)^2}{m-2}m^{g-2}-\frac{4}{m(m-2)}(m-1)^{g+2}\nonumber\\
&\quad&-\frac{(3m-2)(m-1)(m^2+2)}{(m-2)(2m-1)}\left(\frac{m}{m-1}\right)^{g-2}\nonumber\\
&\quad&+\frac{1}{m(m-2)(2m-1)}[(2m^4-7m^3+7m^2-2m)g\nonumber\\
&\quad&+3m^5-11m^4+19m^3-22m^2+16m-4].
\end{eqnarray}
\end{small}

Both Eqs.~(\ref{F2}) and~(\ref{F4}) indicate that for large networks ($g \to\infty$),
\begin{eqnarray}\label{F5}
T_{h f}(g) \sim m^g\,
\end{eqnarray}
implying that $T_{h f}(g)$ behaves as a linear function of the network size $N_g$.

\subsubsection{Dominant scaling of ATT to the trap}

In the above we have found the rigorous solution to the MFPT $T_{h f}(g)$ from the root to a farthest node in $M_g$, and have shown that $T_{h f}(g)$ scales linearly with the network size $N_g$. We emphasize that the analytical computation for the ATT $T_{f}(g)$ to a farthest node in $M_g$ is rather lengthy and awkward. However, it is easy to infer that when $g$ is large enough, the dominant term of $T_{f}(g)$ also increases as a linear function of network size $N_g$, which can be understood from the following heuristic explanation. Note that $M_g$ comprises $m$ subgraphs, each of which is a replica of $M_{g-1}$. For those nodes in the central subgraph, their MFPT to the trap is equal to $T_{h}(g-1)+T_{h f}(g)$, the dominant term of which is $T_{h f}(g)$; for nodes in each of the $m-2$ fringe subgraphs $M_{g-1}^{(x)}$ ($x=2,3,\ldots, m-1$), their MFPT to the trap is identical and equals $T_{p}(g-1)+T_{ph}(g)+T_{h f}(g)$ whose dominating term is also $T_{h f}(g)$. At last, for nodes in $M_{g-1}^{(1)}$ containing the trap, the leading term is less than $T_{h f}(g)$. Therefore, we can conclude that for trapping in  $M_g$ with a trap fixed a farthest node, the leading scaling of ATT is the same as that of $T_{h f}(g)$, both of which behave linearly with network size.

\subsection{Result analysis}

Thus far we have studied three cases of trapping in unweighted scale-free modular networks $M_g$ with the trap positioned at a peripheral node, a neighbor node of the root with the smallest degree, and a farthest node from the root, respectively. We have shown that the ATT in $M_g$ exhibits rich behavior. When the trap is located at a root's neighbor with the smallest degree or a farthest node, the ATT behaves linearly with the network size $N_g$; when a peripheral node is considered as the trap, the ATT scales as a linear function of $N_g$ by a logarithmic correction. In addition, our previous works reported~\cite{ZhLiGoZhGuLi09,ZhYaLi12} that when the trap is fixed at the root, the ATT grows sublinearly with $N_g$. These results show that the diffusion processes in $M_g$ are very efficient, since the ATT increases at most linearly with the network size.

Although for trapping in $M_g$, the ATT to different traps displays distinct scalings with the network size, it is easy to check that for the aforementioned trapping problems in $M_g$ the dominating scaling of ATT grows inversely proportional to the degree of the trap, irrespective of trap's location. For example, for the two cases of trapping problems when the trap is fixed on a peripheral node or a farthest node, the dominating scaling of ATT is identical. Again for instance, the dominant scaling for ATT to the root is $N_g/K_h(g)$~\cite{ZhLiGoZhGuLi09,ZhYaLi12}.
In fact, extensive numerical computations also verify that for all trapping problems in the unweighted scale-free modular networks with a deep trap, as long as the degree of the trap node is identical, the leading behavior for their trapping efficiency is also the same.

It has been reported~\cite{TeBeVo09,LiJuZh12,LiZh13} that for trapping problem in an arbitrary sparse network having $N$ nodes with a trap placed at node $j$, the scaling of the lower bound for ATT $T_j$ varies with the network size $N$ as $T_j \sim N/d_j$, in which $d_j$ denotes the degree of node $j$. For trapping problems in the unweighted scale-free modular networks, this minimal scaling for ATT can all be achieved, wherever the trap is located. In this sense, unweighted scale-free modular networks display the optimal structure for efficient diffusion.

\section{Trapping in weighted modular scale-free networks}

In this section, we study random walks in weighted modular scale-free networks $\bar{M}_g$ with a perfect trap positioned at a certain node. Note, random walks is biased, with the transition probability $p_{ij}$ from node $i$ to node $j$ depending on both the weight $w_{ij}$ of edge linking the two nodes and the strength $s_i$ of node $i$ such that
$p_{ij}=w_{ij}/s_i$. Next, we focus on three cases of trapping problems. In the first case, the trap is placed at the root node; in the second case, the trap is located at a neighboring node of the root having the smallest degree; while in the lase case, the trap is fixed at a farthest node. We will show that for all the trapping issues being studied, the ATT can display different scalings by changing the weight parameter $\omega$.

\subsection{Determination of intermediate variables}

In order to study the ATT to the trap for both cases of trapping problems, we first define some intermediate variables and determine their values. We denote by $\bar{T}_{p h}(g)$ the MFPT for a walker starting from an arbitrary peripheral node of $\bar{M}_g$ to visit the root for the first time and by $\bar{T}_{hp}(g)$ the MFPT spent by a walker initially located at the root to first reach any peripheral node. In Appendix~\ref{Ap2}, we give detailed derivations for $\bar{T}_{p h}(g)$ and $\bar{T}_{h p}(g)$. For $\omega \neq m/(m-1)$,
\begin{eqnarray}\label{H3}
\bar{T}_{p h}(g)&=&\frac{(m-1)^2[(\omega-1)m^2-(5\omega-2)m+2\omega]}{(\omega-1)m^3-\omega m^2}\nonumber\\
&\quad&\times\left[\frac{m}{(m-1)\omega}\right]^{g}+1-\frac{2m-2}{m\omega-m-\omega}
\end{eqnarray}
and
\begin{eqnarray}\label{H4}
\bar{T}_{h p}(g)&=&\frac{(m-1)[(\omega-1)m^2-(5\omega-2)m+2\omega]}{(\omega-1)m^3-\omega m^2}\nonumber\\
&\quad&\times\left[\frac{m}{(m-1)\omega}\right]^{g}+1-\frac{2}{m\omega-m-\omega};
\end{eqnarray}
while for $\omega = m/(m-1)$,
\begin{equation}\label{H5}
\bar{T}_{p h}(g)=\frac{2(m-1)}{m} g + \frac{ m^3 - 5m^2 +7m -2 }{m^2}
\end{equation}
and
\begin{equation}\label{H6}
\bar{T}_{h p}(g)= \frac{2}{m} g + \frac{2 m^2 - 5m + 2 }{m^2}\,.
\end{equation}

For the particular case of $\omega=1$, $\bar{M}_g$ is exactly $M_g$, Eqs.~(\ref{H3}) and (\ref{H4}) are reduced to Eqs.~(\ref{B1}) and (\ref{B2}), respectively. The expressions for $T_{p h}(g)$ and $T_{h p}(g)$ are very useful for the following derivation of explicit solutions for ATT to different targets.

\subsection{Trapping with the trap placed at the root}

After obtaining the intermediate quantities, we are now in a position to consider the trapping process in networks $\bar{M}_g$ with the root being the trap. Our goal in this case is to determine the ATT denoted by $\bar{T}_{h}(g)$, which is the average of the MFPT for a walker originating from a node in $\bar{M}_g$ to first visit the root node over all starting points. In order to find $\bar{T}_{h}(g)$, we introduce another quantity $\bar{T}_{p}(g)$, defined as the ATT with all peripheral nodes being occupied by traps. From the structure of the networks we can easily establish the following recursive relations for $\bar{T}_{h}(g)$ and $\bar{T}_{p}(g)$:
\begin{equation}\label{I1}
\bar{T}_{h}(g)=\frac{1}{m}\bar{T}_{h}(g-1)+\frac{m-1}{m}[\bar{T}_{p}(g-1)+\bar{T}_{p h}(g)]
\end{equation}
and
\begin{equation}\label{I2}
\bar{T}_{p}(g)=\frac{1}{m}[\bar{T}_{h}(g-1)+\bar{T}_{h p}(g)]+\frac{m-1}{m}\bar{T}_{p}(g-1).
\end{equation}
The two terms on the rhs of Eq.~(\ref{I1}) can be elaborated as follows: the first term explains the process that a walker starting from a node in central original sub-network $\bar{M}_{g-1}$ takes $\bar{T}_{h}(g-1)$ steps to reach the root. The second term describes the process that the walker leaving from a node in $\bar{M}_{g-1}^{(i)}$ ($1\leq i<m$) takes $\bar{T}_{p}(g-1)$ steps to reach the peripheral nodes of $\bar{M}_{g-1}^{(i)}$, then takes $\bar{T}_{p h}(g)$ steps to visit the root. Equation~(\ref{I2}) can be explained in a similar way.

After some algebra, Eqs~(\ref{I1}) and (\ref{I2}) can be recast as
\begin{equation}\label{I3}
m\bar{T}_{h}(g)-\bar{T}_{h}(g-1)-(m-1)\bar{T}_{p h}(g)=(m-1)\bar{T}_{p}(g-1)
\end{equation}
and
\begin{equation}\label{I4}
m\bar{T}_{p}(g)-(m-1)\bar{T}_{p}(g-1)-\bar{T}_{h p}(g)=\bar{T}_{h}(g-1).
\end{equation}
From Eq.~(\ref{I3}), we can further obtain
\begin{equation}\label{I5}
m\bar{T}_{h}(g+1)-\bar{T}_{h}(g)-(m-1)\bar{T}_{p h}(g+1)=(m-1)\bar{T}_{p}(g),\nonumber\\
\end{equation}
which, together with Eqs.~(\ref{I3}) and (\ref{I4}), yields
\begin{eqnarray}\label{I6}
\frac{m^2}{m-1}\bar{T}_{h}(g+1)&=&\left(\frac{m}{m-1}+m\right)\bar{T}_{h}(g)+m\bar{T}_{p h}(g+1)\nonumber\\
&\quad&-(m-1)\bar{T}_{p h}(g)+\bar{T}_{h p}(g).
\end{eqnarray}
Considering the initial condition $\bar{T}_{h}(2)=(m-1)(m\omega^2+3m\omega+2\omega+m^2-2m-2)/m^2\omega^2$ and plugging Eqs.~(\ref{H3}) and (\ref{H4}) into Eq.~(\ref{I6}), we can solve inductively Eq.~(\ref{I6}) to obtain the following rigorous expressions:
\begin{widetext}
\begin{eqnarray}\label{I7}
\bar{T}_{h}(g)&=&\frac{2(m-1)^2\omega}{m^2(m\omega-m-\omega)}g
+\frac{(m\omega-m-2\omega)(m^2\omega-m^2-5m\omega+2m+2\omega)(m-1)^3}{(m+\omega-m\omega)^2m^3}\left[\frac{m}{(m-1)\omega}\right]^g\nonumber\\
&\quad&+\frac{(m-2)(m-1)^3\omega^2+(m^2\omega-6m\omega-2m^2+7\omega+5m-2)(m-1)m-2(m-1)\omega}{(m+\omega-m\omega)^2m^2}
\end{eqnarray}
\end{widetext}
and
\begin{eqnarray}\label{I7c}
\bar{T}_{h}(g)&=&\frac{m-1}{m^2} g^2 + \frac{3m^3 - 9m^2 + 8m - 2}{m^3} g \nonumber\\
&\quad&+ \left( m +\frac{16}{m} -\frac{14}{m^2} + \frac{4}{m^3} - 7 \right),
\end{eqnarray}
which hold for $\omega \neq m/(m-1)$ and $\omega = m/(m-1)$, respectively.
Particularly, for $\omega=1$, Eq.~(\ref{I7}) reduces to Eq.~(\ref{B15}).

\begin{figure}
\begin{center}
\includegraphics[width=1.0\linewidth,trim=0 15 0 15]{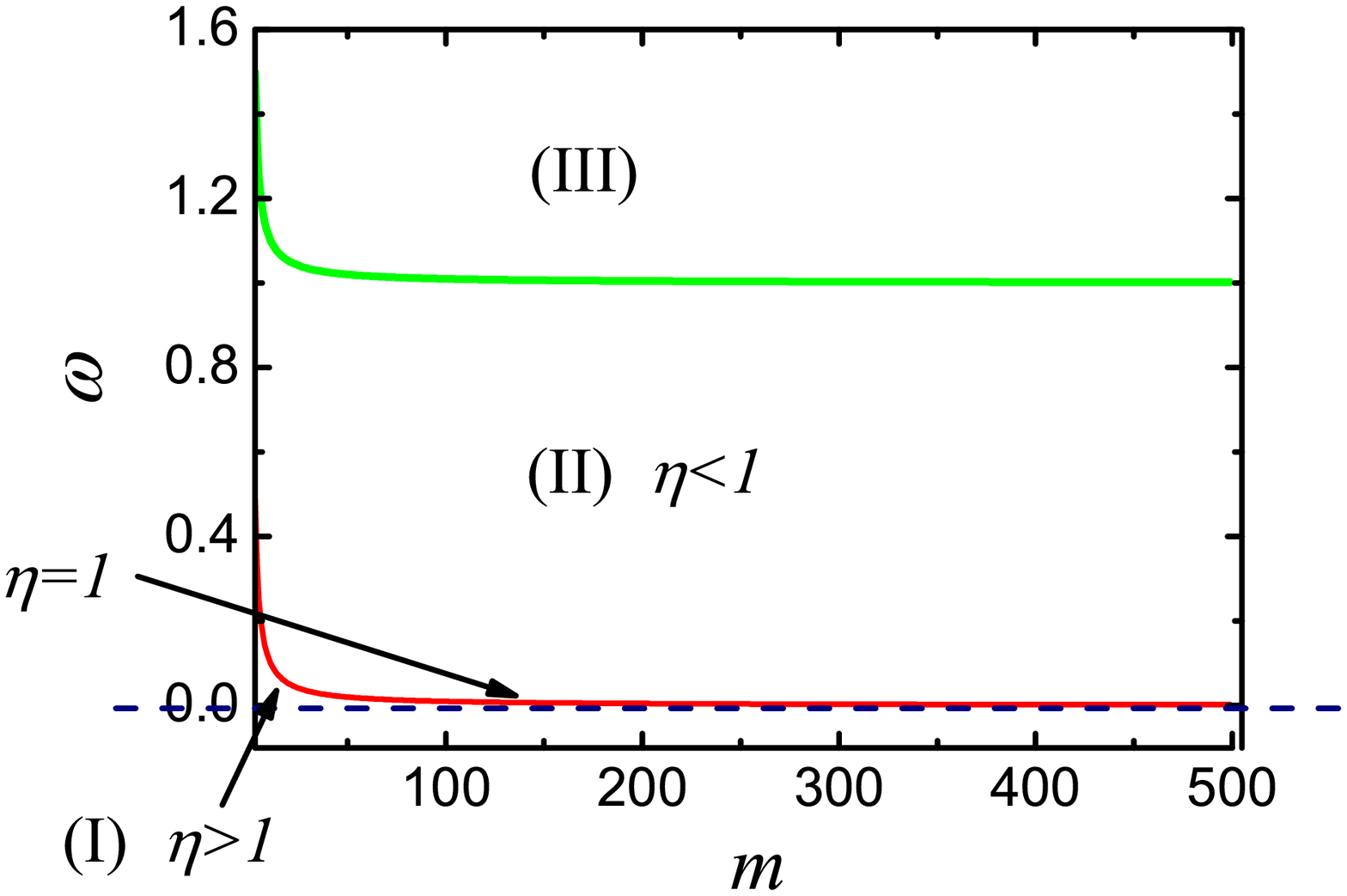}
\end{center}
\caption[kurzform]{(Color online) Regions of the $(m,\omega)$ plane highlighting the behavior of $\bar{T}_{h}(g)$. The regions are
defined as (I) $0<\omega < 1/(m-1)$ where $\bar{T}_{h}(g)\sim (N_{g})^{\eta(\omega)}$ with $\eta(\omega)>1$, (II) $1/(m-1)< \omega < m/(m-1)$ where $\bar{T}_{h}(g)\sim (N_{g})^{\eta(\omega)}$ with $\eta(\omega)<1$, and (III) $\omega>m/(m-1)$ where $\bar{T}_{h}(g)\sim \ln N_{g}$. At
the red boundary  $\omega= 1/(m-1)$ between (I) and (II) and the green boundary  $\omega= m/(m-1)$  between (II) and (III), $\bar{T}_{h}(g)\sim N_{g}$ and $\bar{T}_{h}(g)\sim (\ln N_{g})^{2}$, respectively.} \label{HubZone}
\end{figure}

Equations~(\ref{I7}) and~(\ref{I7c}) show that for trapping in weighted modular scale-free networks $\bar{M}_g$, the ATT to the root depends on the weight parameter $\omega$. When $\omega>m/(m-1)$, the first term on the rhs of Eq.~(\ref{I7}) dominates. In this case, the ATT is proportional to generation $g$, which is a logarithmical function of network size $N_g$, since $N_g=m^g$. When $\omega= m/(m-1)$, From Eq.~(\ref{I7c}) we have $\bar{T}_{h}(g)\sim g^{2}$, implying that $\bar{T}_{h}(g)\sim (\ln N_{g})^{2}$.  When $\omega< m/(m-1)$, the second term on the rhs of Eq.~(\ref{I7}) dominates. In this case the ATT $\bar{T}_{h}(g)$ behaves as a power-law function of $N_{g}$:
\begin{equation}\label{I8}
\bar{T}_{h}(g)\sim (N_g)^{\eta(\omega)}=(N_g)^{1-\ln [\omega(m-1)]/\ln m}\,,
\end{equation}
where the exponent $\eta(\omega)=1-\ln [w(m-1)]/\ln m$ can be less than, equal to, or greater than $1$, dependent on the parameter $\omega$. If $0<\omega < 1/(m-1)$,
$\bar{T}_{h}(g)$ varies superlinearly with $N_{g}$; if $\omega= 1/(m-1)$, $\bar{T}_{h}(g)$ scales linearly with $N_{g}$; and if $1/(m-1)< \omega < m/(m-1)$, $\bar{T}_{h}(g)$ behaves
sublinearly with $N_{g}$. Figure~\ref{HubZone} schematically presents the regions in the $(m, \omega)$ plane, where the networks display different dominant scalings in the ATT $\bar{T}_{h}(g)$.

Therefore, the ATT $\bar{T}_{h}(g)$ to the root exhibits rich behavior by adjusting
$\omega$ to control (change) transition probability from one node to another: when $\omega$ grows from zero to infinite, the
trapping efficiency measured by $\bar{T}_{h}(g)$ covers a range from superlinear dependence (less
efficient trapping) to logarithmical dependence (highly efficient trapping) on the network size $N_{g}$. Since the exponent $\eta(\omega)$ decreases with increasing $\omega$, comparing Eqs.~(\ref{B15}) with Eqs.~(\ref{I7}) and~(\ref{I7c}), we can conclude that for the case $\omega > 1$, $\bar{T}_{h}(g)$ grows slower than $T_{h}(g)$, and for the case $0<\omega<1$, $\bar{T}_{h}(g)$ increases faster than $T_{h}(g)$.

\subsection{Trapping with the trap located at a root's neighbor with the smallest degree}

We here consider trapping in $\bar{M}_g$ with the trap being placed at a neighbor node $x$ of the root, having the smallest degree. Let $\bar{T}_{x}(g)$ be the ATT to the trap node $x$. In order to derive $\bar{T}_{x}(g)$, we first evaluate a related quantity, $\bar{T}_{hx}(g)$, defined as the MFPT from the root to node $x$. By construction, $\bar{T}_{h x}(g)$ satisfies
\begin{eqnarray}\label{J1}
\bar{T}_{h x}(g)&=&\frac{1}{\displaystyle{\sum_{k=1}^{g}}(m-1)^k\omega^k}\Bigg\{\sum_{k=2}^g(m-1)^k\omega^k[1+\bar{T}_{p h}(k)\nonumber\\
&\quad&+\bar{T}_{h x}(g)]+\omega+\omega(m-1)[1+\bar{T}_{y x}(g)]\Bigg\},\nonumber\\
\end{eqnarray}
where $\bar{T}_{y x}(g)$ is the MFPT from a node $y$ ($y \in \Omega$ and $y \neq x$) to node $x$, which obeys \begin{eqnarray}\label{J2}
\bar{T}_{y x}(g)&=&\frac{1}{\omega+m-2}\big\{1+(m-3)[1+\bar{T}_{y x}(g)]\nonumber\\&\quad&+\omega[1+\bar{T}_{h x}(g)]\big\}.
\end{eqnarray}

The three terms on the rhs of Eq.~(\ref{J1}) can be explained as follows. The first term describes the process that the walker starting from the root has a probability of $(m-1)^k\omega^k/S_{h}(g)$ arriving at a local peripheral node in $\mathbb P_k$ ($2\leq k\leq g$) after one step, then takes another $\bar{T}_{p h}(k)$ steps back to the root, from which it continues to make $\bar{T}_{h x}(g)$ jumps reaching the target. The second term explains the fact that with probability $\omega/S_{h}(g)$, the walker only makes one jump to hit the destination. And the third term accounts for the fact that with probability $\omega(m-1)/S_{h}(g)$ the walker first jumps to a node  $y$ in one time step, then take $\bar{T}_{y x}(g)$ more time steps to arrive at node $x$. Analogously, one can account for Eq.~(\ref{J2}).

From Eq.~(\ref{J2}), $\bar{T}_{y x}(g)$ can be expressed in terms of $\bar{T}_{h x}(g)$ as
\begin{eqnarray}\label{J3}
\bar{T}_{y x}(g)=\frac{\omega+m-2+\omega\bar{T}_{h x}(g)}{\omega+1}.
\end{eqnarray}
Inserting Eq.~(\ref{J2}) into Eq.~(\ref{J1}) yields
\begin{eqnarray}\label{J4}
\bar{T}_{h x}(g)&=&\frac{\omega+1}{(m+\omega-1)\omega}\bigg[\sum_{k=2}^{g}(m-1)^k\omega^k\bar{T}_{p h}(k)\nonumber\\
&\quad&+\sum_{k=1}^{g}(m-1)^k\omega^k+\frac{\omega(m-2)(\omega+m-2)}{\omega+1}\bigg].\nonumber\\
\end{eqnarray}
Plugging Eq.~(\ref{H3}) into Eq.~(\ref{J4}), the quantity $\bar{T}_{h x}(g)$ can be evaluated as
\begin{eqnarray}\label{J5}
\bar{T}_{h x}(g)&=&\frac{(m-1)(\omega+1)(m^2\omega-5m\omega-m^2+2m+2\omega)}{(m\omega-m-\omega)(m+\omega-1)m\omega}m^g\nonumber\\
&\quad&+\frac{2(m-1)(\omega+1)}{(m+\omega-1)(m\omega-m-\omega)}[(m-1)\omega]^g\nonumber\\
&\quad&-\frac{m+\omega-2}{\omega}
\end{eqnarray}
and
\begin{eqnarray}\label{J5c}
\bar{T}_{h x}(g)&=&\frac{2 (2m-1) (m -1)}{(m^2 - m + 1)m} g m^g \\ \nonumber
&& + \frac{(m-2) (m-1)^2 (2 m-1)}{(m^2 - m + 1)m^2} m^g \\ \nonumber
&& - \left(m+\frac{2}{m}-2\right)\,,
\end{eqnarray}
both of which are true for $\omega \neq m/(m-1)$ and $\omega = m/(m-1)$, respectively.

Having obtained $\bar{T}_{h x}(g)$, we continue to derive the ATT $\bar{T}_{x}(g)$ to node $x$, which can be computed as
\begin{eqnarray}\label{J6}
\bar{T}_{x}(g)&=&\frac{1}{m^g}\bigg[m^g\bar{T}_{h}(g)-(m-1)\left(1+\frac{m-2}{\omega}\right)\nonumber\\
&\quad&+(m^g-m+1)\bar{T}_{h x}(g)\nonumber\\
&\quad&+(m-2)\frac{\omega\bar{T}_{h x}(g)+m+\omega-2}{\omega+1}\bigg]
\end{eqnarray}
Inserting Eqs.~(\ref{I7}),~(\ref{I7c}),~(\ref{J5}) and~(\ref{J5c}) into Eq.~(\ref{J6}), we obtain the closed-form expressions for $\bar{T}_x(g)$:
\begin{widetext}
\begin{eqnarray}\label{J7}
\bar{T}_{x}(g)&=&\frac{(m-1)(\omega+1)(m^2\omega-5m\omega-m^2+2m+2\omega)}{(m\omega-m-\omega)(m+\omega-1)m\omega}m^g+\frac{2(m-1)(\omega+1)}{(m+\omega-1)(m\omega-m-\omega)}[(m-1)\omega]^g\nonumber\\
&\quad&+\frac{(m\omega-m-2\omega)(m^2\omega-m^2-5m\omega+2m+2\omega)(m-1)^3}{(m+\omega-m\omega)^2m^3}\left[\frac{m}{(m-1)\omega}\right]^g+\frac{2(m-1)^2\omega}{m^2(m\omega-m-\omega)}g\nonumber\\
&\quad&-\frac{2(m-1)}{m\omega-m-\omega}\left[\frac{(m-1)\omega}{m}\right]^g-\frac{3m-2}{m^2}+\frac{(m-1)(5m^2-9m+2)}{m^2(m\omega-m-\omega)}-\frac{2m^2-5m+2}{m\omega}-\frac{2(m-1)}{(m\omega-m-\omega)^2}\nonumber\\
\end{eqnarray}
\end{widetext}
and
\begin{eqnarray}\label{J7c}
\bar{T}_{x}(g)&=&\frac{2(2 m-1)(m-1)}{(m^2 -m + 1)m} g m^g  \\ \nonumber
&& + \frac{(m-2) (m-1)^2 (2 m-1)}{(m^2 -m + 1)m^2} m^g \\ \nonumber
&& + \frac{ m - 1}{m^2} g^2 + \frac{(m-1) (m^2 - 6m + 2)}{m^3} g \\ \nonumber
&& - \frac{(m^2-3 m+3)(m+4)m - 4}{m^3}
\end{eqnarray}
for $\omega \neq m/(m-1)$ and $\omega = m/(m-1)$, respectively.

\begin{figure}
\begin{center}
\includegraphics[width=1.0\linewidth,trim=0 15 0 15]{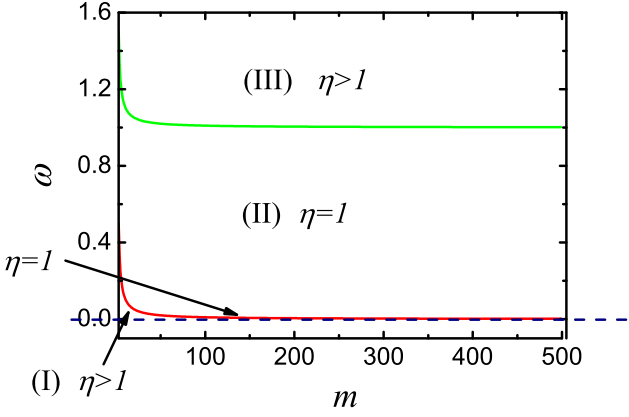}
\end{center}
\caption[kurzform]{(Color online) Regions of the $(m,\omega)$ plane specifying the leading scalings of $\bar{T}_{x}(g)$. The regions are
defined as (I) $0<\omega < 1/(m-1)$ where $\bar{T}_{h}(g)\sim (N_{g})^{\eta(\omega)}$ with $\eta(\omega)>1$, (II) $1/(m-1)\leq \omega < m/(m-1)$ where $\bar{T}_{h}(g)\sim N_{g}$, and (III) $\omega>m/(m-1)$ where $\bar{T}_{h}(g)\sim (N_{g})^{\eta(\omega)}$ with $\eta(\omega)>1$. At
the  boundary line $\omega= m/(m-1)$  between (II) and (III), $\bar{T}_{h}(g)\sim N_{g}\ln N_{g}$.} \label{NeighborZone}
\end{figure}

Similar to the case when the trap is located at the root, Eqs.~(\ref{J7}) and~(\ref{J7c}) show that for trapping in $\bar{M}_g$ with the trap being fixed at a neighboring node $x$ of the root, the ATT to the trap is also determined by the weight parameter $\omega$. When $0<\omega < 1/(m-1)$ the third term on the rhs of Eq.~(\ref{J7}) is the leading one. In this case, the ATT $\bar{T}_{x}(g)$ to the trap varies superlinearly with the network size $N_{g}$. When $\omega=1/(m-1)$, the first and third terms prevail; but when $1/(m-1)<\omega< m/(m-1)$, only the first term dominates. That is, for $1/(m-1)\leq \omega< m/(m-1)$, $\bar{T}_{x}(g)$ grows linearly with $N_{g}$; for $\omega= m/(m-1)$, $\bar{T}_{x}(g)$ scales with $N_{g}$ as $\bar{T}_{x}(g) \sim N_{g}\ln N_{g}$. Finally, when $\omega>m/(m-1)$, the second term on the rhs of Eq.~(\ref{J7}) dominates, implying that $\bar{T}_{x}(g)$ scales superlinearly with $N_{g}$. In Fig.~\ref{NeighborZone}, we present the zones in the $(m, \omega)$ plane, where the scalings of the ATT $\bar{T}_{x}(g)$ are distinct. From Fig.~\ref{NeighborZone}, we can see that different from the case that the root is trap, when one of root's neighbors with the smallest degree is the trap, the ATT is no more a monotonous function of $\omega$.

\subsection{Trapping with the trap positioned at a peripheral node}

For the case that the trap is fixed at one of the $(m-1)^g$ peripheral nodes, let $\bar{T}_{h \mathcal B_0}(g)$ denote the MFPT from the root to the trap, and $\bar{T}_{{\mathcal B}_0}(g)$   stand for the ATT. After lengthy and awkward calculations, we obtain
\begin{widetext}
\begin{eqnarray} \label{HtoB}
\bar{T}_{h{\mathcal B}_0} &=& \frac{(m-2)(\omega-1)}{m(m \omega -m-\omega)} \sum _{j=1}^{g-1} \frac{(m-1)^{g-j}  \omega^{g-1} \left\{ (m^2\omega-m^2-5m \omega+2m+2 \omega)\left[\frac{m}{(m-1) \omega}\right]^g+2 m \omega\right\} }{ \omega^g-\omega^j} \nonumber\\
&&+\frac{(m-1) (m^2\omega-m^2-5m \omega+2m+2 \omega)\left[\frac{m}{(m-1) \omega}\right]^g +m^2 (m \omega-m-\omega+2)}{m^2 (m \omega - m -\omega)} \nonumber\\
&&+\frac{2m\omega (m-2) (\omega-1) \left[(m-1) \omega\right]^g + (m-2) (\omega-1) (m^2 \omega-m^2-5m \omega+2m+ 2 \omega) m^g }{m (m\omega-m-\omega) \left(\omega^{g+1} +m \omega -m -2\omega +1 \right)}\,.
\end{eqnarray}
\end{widetext}
Since the dominant scaling of $\bar{T}_{{\mathcal B}_0}(g)$ is similar to that $\bar{T}_{h \mathcal B_0}(g)$, with the exception of the cases of $\omega= 1/(m-1)$ and $\omega= m/(m-1)$, we can conclude that the leading scaling of ATT $\bar{T}_{{\mathcal B}_0}(g)$ depends on the weight parameter $\omega$ in the following way: When $0<\omega < 1/(m-1)$, $\bar{T}_{{\mathcal B}_0}(g)\sim (N_{g})^{\eta(\omega)}$ with $\eta(\omega)>1$;  when $\omega=1/(m-1)$, $\bar{T}_{{\mathcal B}_0}(g) \sim N_{g}\ln N_{g}$; when $1/(m-1)<\omega< 1$, $\bar{T}_{{\mathcal B}_0}(g) \sim N_{g}$; when $\omega=1$, $\bar{T}_{{\mathcal B}_0}(g) \sim N_{g}/ \ln N_{g}$; when
$1<\omega< m/(m-1)$, $\bar{T}_{{\mathcal B}_0}(g)\sim (N_{g})^{\eta(\omega)}$ with $\eta(\omega)<1$; when
$\omega= m/(m-1)$, $\bar{T}_{{\mathcal B}_0}(g)\sim (N_{g})^{\ln(m-1)/\ln m}\ln N_{g}$; and when
$\omega> m/(m-1)$, $\bar{T}_{{\mathcal B}_0}(g)\sim (N_{g})^{\ln(m-1)/\ln m}$.

Note that for the case that the trap is placed at a farthest node, it is very hard and even impossible to obtain an analytical formula for the ATT. But considering the facts that for each trapping problem studied above, the ATT is approximatively equal to the total strength and the inverse of the trap's strength for most cases of $\omega$, and that the strength of a farthest node is similar to that of a hub's neighbor with the smallest degree, we guess that the leading behavior of the ATT to a farthest node is analogous to that corresponding to $T_x(g)$.

Before closing this section, it is worth mentioning that the above obtained results on the ATT for all the trapping problems addressed in the weighted networks are (at least quantitatively) expected, since $\omega>1$ implies that the root is easier to be reached and vice versa for $\omega<1$.



\section{Conclusion}

We have studied random walks taking place in a family of unweighted or weighted networks in the presence of a deep trap, which exhibit the remarkable scale-free, small-world, and modular characteristics observed for various real-life social and biological networks. For trapping in binary networks, we have presented an extensive analysis on some cases of trapping problems, with the trap being located at a peripheral node, a neighbor node of the root with the least degree, and a farthest node, respectively. For trapping in the proposed weighted networks with a tunable weight parameter, we have also studied three cases of trapping issues with the trap positioned at the root, a neighboring node of the root having the least degree, or a peripheral node.

For most of the trapping problems in the binary or weighted networks, we have deduced explicit solutions to the ATT, as well as their leading behavior. We have demonstrated that for all the trapping problems in binary networks, the dominating scaling for ATT is proportional to the size of the network and the inverse degree of the trap, which is equivalent to the predicted minimal scaling. Thus, the architecture of the networks being studied are optimal, in the sense that it is instrumental to efficient diffusion. Furthermore, we have shown that by varying the weight parameter, the ATT exhibits rich behavior, the leading scaling of which can be a superlinear, linear, sublinear or logarithmical function of the system size. Thus, one can control the trapping process in the weighted scale-free modular networks by adjusting the weight parameter, which can also be applied to tailor the networks to carry out other desirable functions as wanted.

\begin{acknowledgments}
The authors thank Yuan Lin for her assistance in preparing this manuscript. This work was supported by the National Natural Science Foundation of China under Grant Nos. 61074119 and 11275049.
\end{acknowledgments}

\appendix

\section{Derivation of $p_g(i)$ and $h_g(i)$\label{Ap1}}

We first consider the case $0\leq i<g-2$. In this case, the two quantities $p_g(i)$ and $h_g(i)$ satisfy the following relations:
\begin{small}
\begin{eqnarray}\label{G1}
p_g(i)&=&\frac{1}{g-i+m-2}\bigg\{(m-2)[1+p_g(i)]+[1+h_g(i-1)\nonumber\\
&\quad&+p_g(i)]+1+\sum_{k=1}^{g-i-2}[1+T_{h p}(k)+p_g(i)]\bigg\}
\end{eqnarray}
\end{small}
and
\begin{small}
\begin{eqnarray}\label{G2}
h_g(i)&=&\frac{1}{\displaystyle{\sum_{k=1}^{g-i}}(m-1)^k}\bigg\{(m-1)^{g-i}[1+p_g(i-1)+h_g(i)]\nonumber\\
&\quad&+\frac{1}{m-1}(m-1)^{g-i-1}\nonumber\\
&\quad&+(m-2)(m-1)^{g-i-2}[1+T_{p h}(g-i-1)+h_g(i)]\nonumber\\
&\quad&+\sum_{k=1}^{g-i-2}(m-1)^k[1+T_{p h}(k)+h_g(i)]\bigg\}.
\end{eqnarray}
\end{small}
Equation~(\ref{G1}) can be elaborated as follows. Originating from a node in ${\mathcal R}_{g-i}$, the particle can jump to one of the $m-2$ neighboring nodes in ${\mathcal R}_{g-i}$, from which it continues to jump $p_g(i)$ steps to first visit the destination; this is accounted for by the first term on the rhs. Alternatively, the walker can go to a local hub belonging to ${\mathcal H}_{g-(i-1)}$, then proceeds to bounce $h_g(i-1)+p_g(i)$ steps to hit the target. Such a process is depicted by the second term. The third term describes the process that the walker goes directly to the trap. Finally, the last sum term explains the fact that the particle goes to a local hub in ${\mathbb H}_{g-i}^{(1)}$, from which it takes an average time $T_{h,p}(i)+p_g(i)$ to reach the trap. Analogously, we can explain Eq.~(\ref{G2}).

After some algebra, Eqs.~(\ref{G1}) and (\ref{G2}) can be simplified to
\begin{eqnarray}\label{G3}
p_g(i)=h_g(i-1)+\sum_{k=1}^{g-i-2}T_{h p}(k)+g-i+m-2
\end{eqnarray}
and
\begin{small}
\begin{eqnarray}\label{G4}
h_g(i)&=&(m-1)^2p_g(i-1)\nonumber\\
&\quad&+\frac{1}{(m-1)^{g-i-2}}\sum_{k=1}^{g-i-2}(m-1)^k T_{p h}(k)\nonumber\\
&\quad&+(m-2)T_{p  h}(g-i-1)+\frac{1}{(m-1)^{g-i-2}}\sum_{k=1}^{g-i}(m-1)^k.\nonumber\\
\end{eqnarray}
\end{small}
Inserting Eq.~(\ref{G4}) into Eq.~(\ref{G3}) and utilizing the initial condition
\begin{equation}\label{G5}
p_g(0)=(3m-2)(m+1)\left(\frac{m}{m-1}\right)^{g-3}-2m^2+4m-1,
\end{equation}
Eq.~(\ref{G3}) is solved to get
\begin{eqnarray}\label{G6}
p_g(i)&=&(3m-2)\frac{m^{g-1}}{(m-1)^{g-i-2}}-\frac{3m-2}{m}\left(\frac{m}{m-1}\right)^{g-i-2}\nonumber\\
&\quad&-2(m-1)^{i+2}+1,
\end{eqnarray}
plugging which into Eq.~(\ref{G4}) yields
\begin{eqnarray}\label{G7}
h_g(i)&=&(3m-2)\frac{m^{g-1}}{(m-1)^{g-i-3}}-(3m-2)\left(\frac{m}{m-1}\right)^{g-i-3}\nonumber\\
&\quad&-2(m-1)^{i+3}+2m-3.
\end{eqnarray}

For the case $i=g-2$, the quantities of $p_g(g-2)$ and $h_g(g-2)$ are given by
\begin{eqnarray}\label{G8}
p_g(g-2)=(3m-2)m^{g-1}-2(m-1)^g-1
\end{eqnarray}
and
\begin{eqnarray}\label{G9}
h_g(g-2)&=&2(3m-2)(m-1)m^{g-2}-\frac{2}{m}(3m-2)(m-1)\nonumber\\
&\quad&-\frac{4}{m}(m-1)^{g+1}+\frac{1}{m}(5m-4)(m-1)\,,
\end{eqnarray}
respectively.

\section{Derivation of $\bar{T}_{p h}(g)$ and $\bar{T}_{h p}(g)$\label{Ap2}}

By construction, for any $g>1$, the two quantities $\bar{T}_{p h}(g)$ and $\bar{T}_{h p}(g)$ satisfy the following two recursive relations:
\begin{eqnarray}\label{K1}
\bar{T}_{p h}(g)&=&\frac{1}{m-2+\displaystyle{\sum_{k=1}^{g}}\omega^k}\Bigg\{(m-2)[1+\bar{T}_{p h}(g)]\nonumber\\
&\quad&+\sum_{k=1}^{g-1}[1+\bar{T}_{h p}(k)+\bar{T}_{p h}(g)]\omega^k+\omega^g\Bigg\}
\end{eqnarray}
and
\begin{eqnarray}\label{K2}
\bar{T}_{h p}(g)&=&\frac{1}{\displaystyle{\sum_{k=1}^{g}}(m-1)^k\omega^k}\Bigg \{(m-1)^g\omega^g\nonumber\\
&\quad&+\sum_{k=1}^{g-1}[1+\bar{T}_{p h}(k)+\bar{T}_{h p}(g)](m-1)^k\omega^k\Bigg \}.\nonumber\\
\end{eqnarray}
The three terms on the rhs of Eq.~(\ref{K1}) can be explained as follows. For any peripheral node, its strength is $S_{p}(g)=m-2+\sum_{k=1}^{g}\omega^k$, as shown in Eq.~(\ref{A14}). The first term describes the process that with probability $(m-2)/S_{p}(g)$ a particle leaving from a peripheral makes a jump to another peripheral node, and then spends $\bar{T}_{p h}(g)$ more time steps to visit the root. In addition, with probability $\omega^k/S_{p}(g)$ the walker may first hop to a local hub in $\mathbb H_k$, then takes $\bar{T}_{h p}(k)$ time steps back to one of the peripheral nodes, and proceeds to jump $\bar{T}_{p h}(g)$ steps to arrive at the target. This process is accounted for by the second term. Finally, the third term is based on the fact that with probability $\omega^g/S_{p}(g)$ the walker makes only a jump  to first reach the root. Analogously, we can elaborate the two terms on the rhs of Eq.~(\ref{K2}).

After some algebra, Eqs.~(\ref{K1}) and (\ref{K2}) can be simplified to
\begin{small}
\begin{eqnarray}\label{K3}
(m-1)^g\omega^g\bar{T}_{h p}(g)&=&\sum_{k=1}^g(m-1)^k\omega^k+\sum_{k=1}^{g-1}(m-1)^k\omega^k\bar{T}_{p h}(k)\nonumber\\
\end{eqnarray}
\end{small}
and
\begin{eqnarray}\label{K4}
\omega^g\bar{T}_{p h}(g)&=&(m-2)+\sum_{k=1}^{g}\omega^k+\sum_{k=1}^{g-1}\omega^k\bar{T}_{h p}(k).
\end{eqnarray}
From Eq.~(\ref{K3}), we can further get
\begin{eqnarray}\label{K5}
(m-1)^{g+1}\omega^{g+1}\bar{T}_{h p}(g+1)&=&\sum_{k=1}^{g+1}(m-1)^{k}\omega^k\nonumber\\
&\quad&+\sum_{k=1}^{g}(m-1)^k\omega^k\bar{T}_{p h}(k).\nonumber\\
\end{eqnarray}
Combining Eqs.~(\ref{K3}) and (\ref{K5}) yields
\begin{eqnarray}\label{K6}
(m-1)\omega\bar{T}_{h p}(g+1)-\bar{T}_{h p}(g)=(m-1)\omega+\bar{T}_{p h}(g).
\end{eqnarray}
Similarly, we can obtain
\begin{eqnarray}\label{K7}
\omega\bar{T}_{ph}(g+1)-\bar{T}_{p h}(g)=\omega+\bar{T}_{h p}(g),
\end{eqnarray}
which, together with Eq.~(\ref{K6}), leads to
\begin{eqnarray}\label{K8}
(m-1)\omega\bar{T}_{h p}(g+2)-m\bar{T}_{h p}(g+1)=(m-1)\omega-m+2.
\end{eqnarray}
Considering $\bar{T}_{h,p}(2)=1+(2\omega+m-2)/(m-1)\omega^2$, Eq.~(\ref{K8}) can be solved to yield the analytical expression for $\bar{T}_{h p}(g)$ as
\begin{eqnarray}\label{K9}
\bar{T}_{hp}(g)&=&\frac{(m-1)[(\omega-1)m^2-(5\omega-2)m+2\omega]}{(\omega-1)m^3-\omega m^2}\nonumber\\
&\quad&\times \left[\frac{m}{(m-1)\omega}\right]^{g}+1-\frac{2}{m\omega-m-\omega}
\end{eqnarray}
and
\begin{equation}\label{K9c}
\bar{T}_{h p}(g)= \frac{2}{m} g + \frac{2 m^2 - 5m + 2 }{m^2}
\end{equation}
for $\omega \neq m/(m-1)$ and $\omega = m/(m-1)$, respectively.
Instituting Eqs.~(\ref{K9}) and~(\ref{K9c}) into Eq.~(\ref{K7}) and using the initial condition $\bar{T}_{p h}(2)=1+(2\omega+m-2)/\omega^2$, Eq.~(\ref{K7}) is solved to get
\begin{eqnarray}\label{K10}
\bar{T}_{p h}(g)&=&\frac{(m-1)^2[(\omega-1)m^2-(5\omega-2)m+2\omega]}{(\omega-1)m^3-\omega m^2}\nonumber\\
&\quad&\times \left[\frac{m}{(m-1)\omega}\right]^{g}+1-\frac{2m-2}{m\omega-m-\omega}
\end{eqnarray}
for $\omega \neq m/(m-1)$ and
\begin{equation}\label{K10c}
\bar{T}_{p h}(g)=\frac{2(m-1)}{m} g + \frac{ m^3 - 5m^2 +7m -2 }{m^2}\,
\end{equation}
for $\omega = m/(m-1)$.



\nocite{*}

\begin{references}


\bibitem{Mo69}
E. W. Montroll,
J. Math. Phys. {\bf 10}, 753 (1969).

\bibitem{BaKlKo97}
A. Bar-Haim, J. Klafter, and R. Kopelman,
J. Am. Chem. Soc. {\bf 119}, 6197  (1997).

\bibitem{BaKl98}
A. Bar-Haim and J. Klafter,
J. Phys. Chem. B {\bf 102}, 1662  (1998).

\bibitem{BaKl98JOL}
A. Bar-Haim and J. Klafter,
J. Lumin. {\bf 76-77}, 197 (1998).


\bibitem{HwLeKa12}
S. Hwang, D.-S. Lee, and B. Kahng,
Phys. Rev. Lett. {\bf 109}, 088701 (2012).

\bibitem{HwLeKa12E}
S. Hwang, D.-S. Lee, and B. Kahng,
Phys. Rev. E {\bf 85}, 046110 (2012).

\bibitem{SoMaBl97}
I. M. Sokolov, J. Mai, and A. Blumen,
Phys. Rev. Lett. {\bf 79}, 857 (1997).

\bibitem{BlZu81}
A. Blumen and G. Zumofen,
J. Chem. Phys. {\bf 75}, 892 (1981).

\bibitem{MuBlAmGiReWe07}
O. M\"{u}lken, A. Blumen, T. Amthor, C. Giese, M. Reetz-Lamour,and M. Weidemuller
Phys. Rev. Lett. {\bf 99}, 090601 (2007).

\bibitem{AgBlMu10IJBC}
E. Agliari, A. Blumen, and O. M\"{u}lken,
Int. J. Bifurcation Chaos {\bf 82}, 012305 (2010).

\bibitem{Ag11}
E. Agliari,
Physica A {\bf 390}, 1853 (2011).   

\bibitem{MuBl11}
O. M\"{u}lken and A. Blumen,
Phys. Rep. {\bf 502}, 37 (2011).
\bibitem{Re01}
S. Redner, \emph{A Guide to First-Passage Processes} (Cambridge
University Press, Cambridge, 2001).

\bibitem{NoRi04}
J. D. Noh and H. Rieger,
%
%
Phys. Rev. Lett. {\bf 92}, 118701 (2004).

\bibitem{BeCoMo05}
O. B\'enichou, M. Coppey, and  M. Moreau,
Phys. Rev. Lett. {\bf 95}, 260601 (2005).

\bibitem{CoBeKl07}
S. Condamin, O. B\'enichou, and J. Klafter,
Phys. Rev. Lett. {\bf 98}, 250602 (2007).

\bibitem{CoBeMo07}
S. Condamin, O. B\'enichou, and M. Moreau,
Phys. Rev. E
{\bf 75}, 021111 (2007).

\bibitem{CoBeTeVoKl07}
S. Condamin, O. B\'enichou, V. Tejedor, R. Voituriez, and J.
Klafter, Nature (London) {\bf 450}, 77 (2007).


\bibitem{GLKo05}
R. A. Garza-L\'opez and J. J. Kozak,
Chem. Phys. Lett. {\bf 406}, 38 (2005).

\bibitem{GLLiYoEvKo06}
R. A. Garza-L\'opez, A. Linares, A. Yoo, G. Evans, and J. J. Kozak,
Chem. Phys. Lett. {\bf 421}, 287 (2006).


\bibitem{KaBa02PRE}
J. J. Kozak and V. Balakrishnan,
Phys. Rev. E {\bf 65}, 021105
(2002).

\bibitem{BeTuKo10}
J. L. Bentz, J. W. Turner, and J. J. Kozak,
Phys. Rev. E {\bf 82}, 011137 (2010).

\bibitem{KaBa02IJBC}
J. J. Kozak and V. Balakrishnan,
Int. J. Bifurcation Chaos {\bf 12}, 2379 (2002).

\bibitem{KaRe89}
B. Kahng and S. Redner,
J. Phys. A: Math. Gen. {\bf 22}, 887 (1989).

\bibitem{Ag08}
E. Agliari,
Phys. Rev. E {\bf 77}, 011128 (2008).

\bibitem{HaRo08}
C. P. Haynes and A. P. Roberts,
Phys. Rev. E {\bf 78}, 041111
(2008).

\bibitem{LiWuZh10}
Y. Lin, B. Wu, and Z. Z. Zhang,
Phys. Rev. E {\bf 82}, 031140 (2010).

\bibitem{ZhWuCh11}
Z. Z. Zhang, B. Wu, and G. R. Chen,
EPL {\bf 96}, 40009 (2011).

\bibitem{WuZh13}
B. Wu  and Z. Z. Zhang,
J. Chem. Phys. {\bf 139}, 024106 (2013).

\bibitem{BeHoKo03}
J. L. Bentz, F. N. Hosseini, and J. J. Kozak,
Chem. Phys. Lett. {\bf 370}, 319  (2003).

\bibitem{BeKo06}
J. L. Bentz and J. J. Kozak,
J. Lumin. {\bf 121}, 62 (2006).


\bibitem{WuLiZhCh12}
B. Wu, Y. Lin, Z. Z. Zhang, and G. R. Chen,
J. Chem. Phys. {\bf 137}, 044903 (2012).


\bibitem{LiZh13JCP}
Y. Lin and Z. Z. Zhang,
J. Chem. Phys. {\bf 138}, 094905 (2013).

\bibitem{KiCaHaAr08}
A. Kittas, S. Carmi, S. Havlin, and P. Argyrakis,
EPL {\bf 84}, 40008 (2008).

\bibitem{ZhQiZhXiGu09}
Z. Z. Zhang, Y. Qi, S. G. Zhou, W. L. Xie, and J. H. Guan,
Phys. Rev. E {\bf 79}, 021127 (2009).

\bibitem{ZhGuXiQiZh09}
Z. Z. Zhang, J. H. Guan, W. L. Xie, Y. Qi, and S. G. Zhou, EPL, {\bf
86}, 10006 (2009).


\bibitem{ZhXiZhGaGu09}
Z. Z. Zhang, W. L. Xie, S. G. Zhou, S. Y. Gao, and J. H. Guan, EPL {\bf 88}, 10001 (2009).

\bibitem{ZhXiZhLiGu09}
Z. Z. Zhang, W. L. Xie, S. G. Zhou, M. Li, and J. H. Guan,
Phys. Rev. E {\bf 80}, 061111 (2009).

\bibitem{ZhYaGa11}
Z. Z. Zhang, Y. H. Yang, and S. Y. Gao, Eur. Phys. J. B {\bf 84}, 331 (2011).

\bibitem{TeBeVo09}
V. Tejedor, O. B\'enichou, and R. Voituriez,
Phys. Rev. E {\bf 80}, 065104(R) (2009).

\bibitem{LiJuZh12}
Y. Lin, A. Julaiti, Z. Z. Zhang,
J. Chem. Phys. {\bf 137}, 124104 (2012).

\bibitem{LiZh13}
Y. Lin and Z. Z. Zhang,
Phys. Rev. E {\bf 87}, 062140 (2013).

\bibitem{BaKl98JPC}
A. Bar-Haim and J. Klafter,
J. Chem. Phys. {\bf 109}, 5187  (1998).


\bibitem{LiSlBa11}
Y.-Y. Liu, J.-J. Slotine, and A.-L. Barab\'asi,
Nature (London) {\bf 473}, 167 (2011).

\bibitem{YaReLaLaLi12}
G. Yan, J. Ren, Y.-C. Lai, C.-H. Lai, and B. W. Li,
Phys. Rev. Lett. {\bf 108}, 218703 (2012).

\bibitem{WaNiLaGr12}
W. X. Wang, X. Ni, Y. C. Lai, and C. Grebogi,
Phys. Rev. E {\bf 85}, 026115 (2012).

\bibitem{LiSlBa13}
Y.-Y. Liu, J.-J. Slotine, and A.-L. Barab\'asi,
Proc. Natl. Acad. Sci. U.S.A. {\bf 110}, 2460 (2013).

\bibitem{YuZhDiWaLa13}
Z. Z. Yuan, C. Zhao, Z. R. Di, W. X. Wang, and Y. C. Lai,
Nat. Commun. {\bf 4}, 2447 (2013).


\bibitem{KoShShTaXuMoBaKl97}
R. Kopelman, M. Shortreed, Z. Y. Shi, W. Tan, Z. Xu, J. S. Moore, A.
Bar-Haim, and J. Klafter, Phys. Rev. Lett. {\bf  78}, 1239 (1997).





\bibitem{RaSoMoOlBa02}
E. Ravasz, A. L. Somera, D. A. Mongru. Z. N. Oltvai, and A.-L.
Barab\'asi, Science {\bf 297}, 1551 (2002).

\bibitem{RaBa03}
E. Ravasz and A.-L. Barab\'asi, Phys. Rev. E {\bf 67}, 026112
(2003).


\bibitem{BaRaVi01}
A.-L. Barab\'asi, E. Ravasz, and T. Vicsek,
Physica A  {\bf 299}, 559 (2001).

\bibitem{IgYa05}
K. Iguchi and H. Yamada, Phys. Rev. E {\bf 71}, 036144 (2005).

\bibitem{ZhLiGaZhGu09}
Z. Z. Zhang, Y. Lin, S. Y. Gao, S. G. Zhou, and J. H. Guan,
J. Stat. Mech. (2009) P10022.

\bibitem{AgBu09}
E. Agliari and R. Burioni,
Phys. Rev. E {\bf 80}, 031125 (2009).

\bibitem{AgBuMa10}
E. Agliari, R. Burioni, and A. Manzotti,
Phys. Rev. E {\bf 82}, 011118 (2010).

\bibitem{MeAgBeVo12}
B. Meyer, E. Agliari, O. B\'enichou, and R. Voituriez,
Phys. Rev. E {\bf 85}, 026113 (2012).

\bibitem{YaZh13}
Y. H. Yang,  and Z. Z. Zhang,
J. Chem. Phys. {\bf 138}, 034101 (2013).

\bibitem{No03}
J. D. Noh, Phys. Rev. E {\bf 67}, 045103(R) (2003).

\bibitem{NoRi04a}
J. D. Noh and H. Rieger, Phys. Rev. E {\bf 69}, 036111 (2004).


\bibitem{ZhYaLi12}
Z. Z. Zhang, Y. H. Yang, and Y. Lin,
Phys. Rev. E {\bf 85}, 011106 (2012).


\bibitem{BaAl99}
A.-L. Barab\'asi and R. Albert,
Science {\bf 286}, 509 (1999).

\bibitem{WaSt98}
D. J. Watts and H. Strogatz,
Nature (London) {\bf 393}, 440 (1998).

\bibitem{ZhLi10}
Z. Z. Zhang and Y. Lin,
J. Stat. Mech. (2010) P12017.

\bibitem{GiNe02}
M. Girvan and M. E. J. Newman, Proc. Natl. Acad. Sci. U.S.A. {\bf
99}, 7821 (2002).


\bibitem{PaDeFaVi05}
G. Palla, I. Der\'enyi, I. Farkas, and T. Vicsek, Nature (London)
{\bf 435}, 814 (2005).

\bibitem{Ne06}
M. E. J. Newman, Proc. Natl. Acad. Sci. U.S.A. {\bf 103}, 8577
(2006).

\bibitem{Fo10}
S. Fortunato, Phys. Rep. {\bf 486}, 75 (2010).

\bibitem{BaBaVe04}
A. Barrat, M. Barth\'elemy, and A. Vespignani,
Phys. Rev. Lett. {\bf 92}, 228701 (2004).

\bibitem{ZhLiGoZhGuLi09}
Z. Z. Zhang, Y. Lin, S. Y. Gao, S. G. Zhou, J. H. Guan, and M. Li,
Phys. Rev. E {\bf 80}, 051120 (2009).




\end{references}
\end{document}